\documentclass[twocolumn,preprintnumbers,amsmath,amssymb,aps,pre,reprint,longbibliography]{revtex4-2}
\usepackage{graphicx}
\usepackage{xcolor}
\begin{document}

\title{Laning Transitions in Pattern Forming Driven Binary Systems with Competing Interactions}
\author{
C. Reichhardt and C. J. O. Reichhardt 
} 
\affiliation{
Theoretical Division and Center for Nonlinear Studies,
Los Alamos National Laboratory, Los Alamos, New Mexico 87545, USA
}

\date{\today}

\begin{abstract}
  A binary system of particles that move in opposite directions under an applied field can exhibit disordered states as well as laned states where the particles organize into oppositely moving high-mobility lanes to reduce collisions. Previous studies of laning transitions generally focused on particles with purely repulsive interactions. Here, we examine laning transitions for oppositely moving pattern-forming systems of particles with competing attractive and repulsive interactions, which in equilibrium form crystal, stripe, and bubble states. In addition to multiple types of laned states, we find jammed crystals, stripes, and bubbles, and generally observe a much richer variety of phases compared to the purely repulsive system. In the stripe forming regime, the system can dynamically reorder into oppositely moving stripes that are aligned in the direction of the drive. The bubble phase can produce strongly polarized jammed states of elongated bubbles where particles in the individual bubbles segregate to opposite sides of the bubbles. We also find disordered states, segregated laned bubble states where the bubbles pass each other in lanes, and segregated bubbles that move through one another. In the compact bubble regime, we obtain a plastic bubble state in which the oppositely driven particles remain trapped in the bubbles but the bubbles move past each other at a slow velocity due to a net imbalance in the bubble population. At higher drives, individual particles begin to jump from bubble to bubble. We show that the different phases and the transitions between them produce signatures in the velocity-force and differential mobility curves. We demonstrate that the critical force for escaping from the jammed state is nonmonotonic, with stripes exhibiting the lowest unjamming force.
\end{abstract}

\maketitle

\section{Introduction}

In a binary assembly of repulsively interacting particles, such as Yukawa particles, subjected to a drive that causes the particles to 
move in opposite directions,
a variety of nonequilibrium phases can appear.
At lower drives, these phases are fluid-like, but for higher drives
there can be a dynamically phase-separated laned state
in which the particles segregate
into extended lanes that move past each other 
\cite{Helbing00,Dzubiella02a,Chakrabarti04,Glanz12,Ikeda12,Klymko16,Poncet17,Isele23,Yu24}.
This type of 
laning has been studied for colloidal particles
\cite{Leunissen05,Rex07,Lowen10,Vissers11a,Geigeneind20},
dusty plasmas \cite{Sutterlin09},
multi-component magnetic skyrmion systems \cite{Vizarim25},
and driven hard disks \cite{Reichhardt18}.
Laning dynamics can also appear
in models of pedestrian flows  \cite{Feliciani16,Bacik23}
and binary driven active matter systems \cite{Bain17,Reichhardt18b,Khelfa22}.
Under certain conditions, dense clusters can span
the system at lower drives, leading to the formation of a
jammed state \cite{Reichhardt18}
or the appearance of intermittent jamming in which
isolated groups of particles cannot get past one
another \cite{Glanz16,Geigeneind20,Yu24}. 
Oppositely driven binary hard disk systems can also exhibit
jammed, disordered, and fully phase-separated laned states,
and the transitions between these states produce
jumps in the velocity-force curves \cite{Reichhardt18}.
Binary active disk systems in which 
the two species are driven in opposite directions
also produce disordered laned states as
well as motility-induced phase-separated states that
coexist with laning \cite{Reichhardt18}.
Laning transitions can also be induced
with ac driving
\cite{Wysocki09,Li21}.
When an additional chirality is introduced, such that 
particles move preferentially to one side when interacting with
other particles, there can be transitions to tilted lane
states. Effects of this type have been observed
for flows of pedestrians with chiral interaction rules \cite{Bacik23}
and in binary driven skyrmion systems
where the chirality arises from a Magnus force \cite{Reichhardt19a,Vizarim25}. 

Most laning studies have been performed for
particles with purely repulsive interactions,
such as screened Coulomb interactions or hard disks,
but there have also been studies in binary systems with
longer-range attraction, such as Lennard-Jones interactions,
where jammed, laned, disordered, and partially laned states
can arise as a function of drive and particle density  \cite{Wachtler16}.  
Laning has also been studied in binary driven dipolar systems,
where a correlation appears between the chaining from the
dipolar interactions and dynamically induced laning  \cite{Kogler15}. 
An open question is what are the dynamic phases and laning states for
a binary assembly of particles with more complex interactions that are
driven in opposite directions.
These interactions could include competing short-range attraction and long-range repulsion (SALR), which leads to the formation
of a variety of different mesophases in the absence of driving,
such as anisotropic crystals, 
bubbles, stripes, and void lattices
\cite{Seul95,Stoycheva00,Reichhardt03,Reichhardt04,Mossa04,Sciortino04,Nelissen05,Liu08,Reichhardt10,McDermott14,Liu19,Hooshanginejad24}. 
Similar mesophase pattern can arise even for purely repulsive interaction
potentials as long as
the interaction potential is composed of
two steps or has two distinct length scales
\cite{Jagla98,Malescio03,Glaser07}. 
SALR interactions can appear in soft matter and biological systems 
\cite{CostaCampos13,AlHarraq22,Hooshanginejad24}.
In addition, a variety of hard condensed matter systems
can be described in terms of particles with competing interactions or
with multiple length scale interactions.
For example, bubble and stripe states are predicted to 
form in electron liquid crystals
\cite{Fogler96,Moessner96,Cooper99,Fradkin99,Gores07,Friess18}
and multi-component superconducting vortex systems
\cite{Xu21,Komendova12,Varney13,Sellin13,Brems22}. 

In this work, we consider laning transitions in an oppositely driven binary assembly of particles with SALR
interactions composed of both long-range and very short-range Coulomb
repulsion and an
intermediate-range attraction.
For a fixed ratio of attraction to repulsion, and in the absence of driving,
the system forms bubbles, stripes, a void lattice,
and finally a dense uniform crystal as a function of increasing
particle density.
For increasing attraction and fixed density,
we find a uniform crystal, a stripe phase, and finally a bubble phase,
where the bubbles decrease in size as the attractive term
becomes larger
\cite{Reichhardt04,Reichhardt10,Nelissen05,Liu08,McDermott16}.
This same model has also been studied with the addition of
a periodic quasi-one-dimensional substrate \cite{Reichhardt24}
or ratchet potential  \cite{Reichhardt24a},
which introduces additional length scales.
In previous work, single species SALR particle assemblies subjected to
uniform dc driving over
random or periodic substrates exhibited
pinned, disordered flow, and dynamically ordered states
\cite{Reichhardt03,Reichhardt24}. 
In this work, we focus on
the dynamical evolution of the patterns when half of the particles are
driven in a direction opposite to that of the other half, and compare
it to the laning transitions found
in purely repulsive systems.
A much richer variety of states arises
in the SALR system.
At zero attraction, we find phases similar to those produced by repulsive
Yukawa particles.
For finite competing interactions, a number of dynamical states appear in
the stripe and bubble regimes,
including disordered flow, fully segregated lanes,
and mixed laning states.
We find that driving can disorder a jammed stripe phase
and cause a dynamical reorientation into stripes aligned with the drive,
but with laning occurring within each stripe.
For bubbles, even in the jammed phase we observe a rearrangement of the
particles in the bubbles, which segregate to form a polarized bubble state.
In the bubble regime we also find a disordered fluid state,
a segregated bubble state with lanes of bubbles,
and segregated states in which the bubbles move through each other.
The different phases and the transitions between them
produce signatures in
the velocity versus drive and
differential mobility versus drive curves,
as well as changes in the density of topological defects and the flow patterns.

\section{Simulation}

We model a two-dimensional system of size $L \times L$
with periodic boundary conditions in the $x$ and $y$ directions
and with $L = 36$.
The system contains
$N$ particles that have a SALR interaction potential with
a long-range repulsive Coulomb term and a short-range attractive term, given by:
\begin{equation} V(R_{ij}) = \frac{1}{R_{ij}} - B\exp(-\kappa R_{ij}) \ . \end{equation}
Here, $R_{ij}=|{\bf R}_i-{\bf R}_j|$ is the distance between particles $i$ and $j$.
The first term is the long-range repulsion, which favors the formation of a triangular lattice, and the second term is the attraction, which can be varied
by changing the strength of the attraction coefficient $B$
and the inverse correlation length $\kappa$.
Unless otherwise noted, we set $\kappa=1.0$.
The particle density is $\rho=N/L^2$.
We initialize the system 
by placing the particles in a triangular lattice and allowing them to
relax into
crystal, stripe, or bubble phases depending on the values of
$\rho$, $B$, and $\kappa$,
as studied in previous work \cite{Reichhardt10,Reichhardt24}.

The particle dynamics evolves
via integrating the following overdamped equation:
\begin{equation} \eta \frac{d {\bf R}_{i}}{dt} = -\sum^{N}_{j \neq i} \nabla V(R_{ij}) + {\bf F}_{\rm DC}^i . \end{equation}
We set the damping coefficient to 
$\eta=1.0$.
The driving force ${\bf F}_{\rm DC}^i = \sigma_{i}F_D{\hat x}$,
where $\sigma_i = +1$ for half of the particles and
$\sigma_i = -1$ for the other half.
In most of the simulations reported here,
we start with $F_D=0$ and gradually increase
$F_D$ by an increment of $\Delta F_D=0.005$ every $10^4$
simulation time steps. When the particle density is low, we consider
smaller values of $\Delta F_D$ and longer time averages.
For other simulations, 
after initializing the system we apply a constant $F_D$.
We measure the time-averaged velocity for only the $\sigma = +1$ particles,
$\langle V \rangle = (2/N)\sum^{N}_{i=1}\delta(\sigma_{i} -1)({\bf v}_{i}\cdot {\hat {\bf x}})$,
where ${\bf v}_{i}$ is the velocity of particle $i$.
When the ratio of $\sigma_i=+1$ and $\sigma_i=-1$ particles is 50:50,
as we consider here,
the corresponding velocity response of the $\sigma = -1$ particles is the
same as that of the $\sigma=+1$ particles but reversed in sign.
We note that for other ratios,
the velocity-force curves can have
different slopes for the different particle species.

\section{Results}

\begin{figure}
\includegraphics[width=\columnwidth]{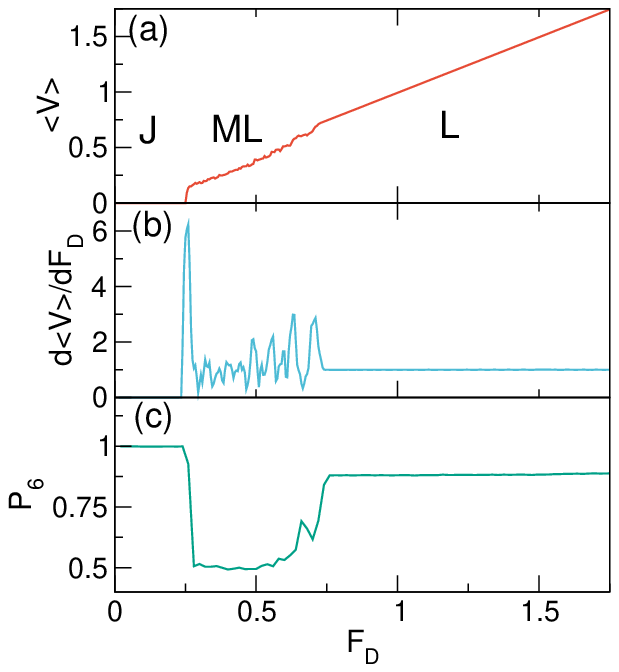}
\caption{(a) Velocity $\langle V \rangle$ vs drive $F_D$
for oppositely driven SALR particles
at a density of $\rho = 0.441$ with $B = 0.0$ in the
purely repulsive Coulomb interaction regime.
Region J is the jammed phase shown in Fig.~\ref{fig:2}(a),
ML is the moving liquid phase shown in Fig.~\ref{fig:2}(b),
and L is the laned phase shown in Fig.~\ref{fig:2}(c).
(b) The $d\langle V \rangle/dF_D$ vs $F_D$ curve more clearly shows the
transitions among phases J, ML, and L.
(c) The fraction of sixfold-coordinated particles, $P_6$, vs
$F_D$ also shows signatures of the three phases.
}
\label{fig:1}
\end{figure}

\begin{figure}
\includegraphics[width=\columnwidth]{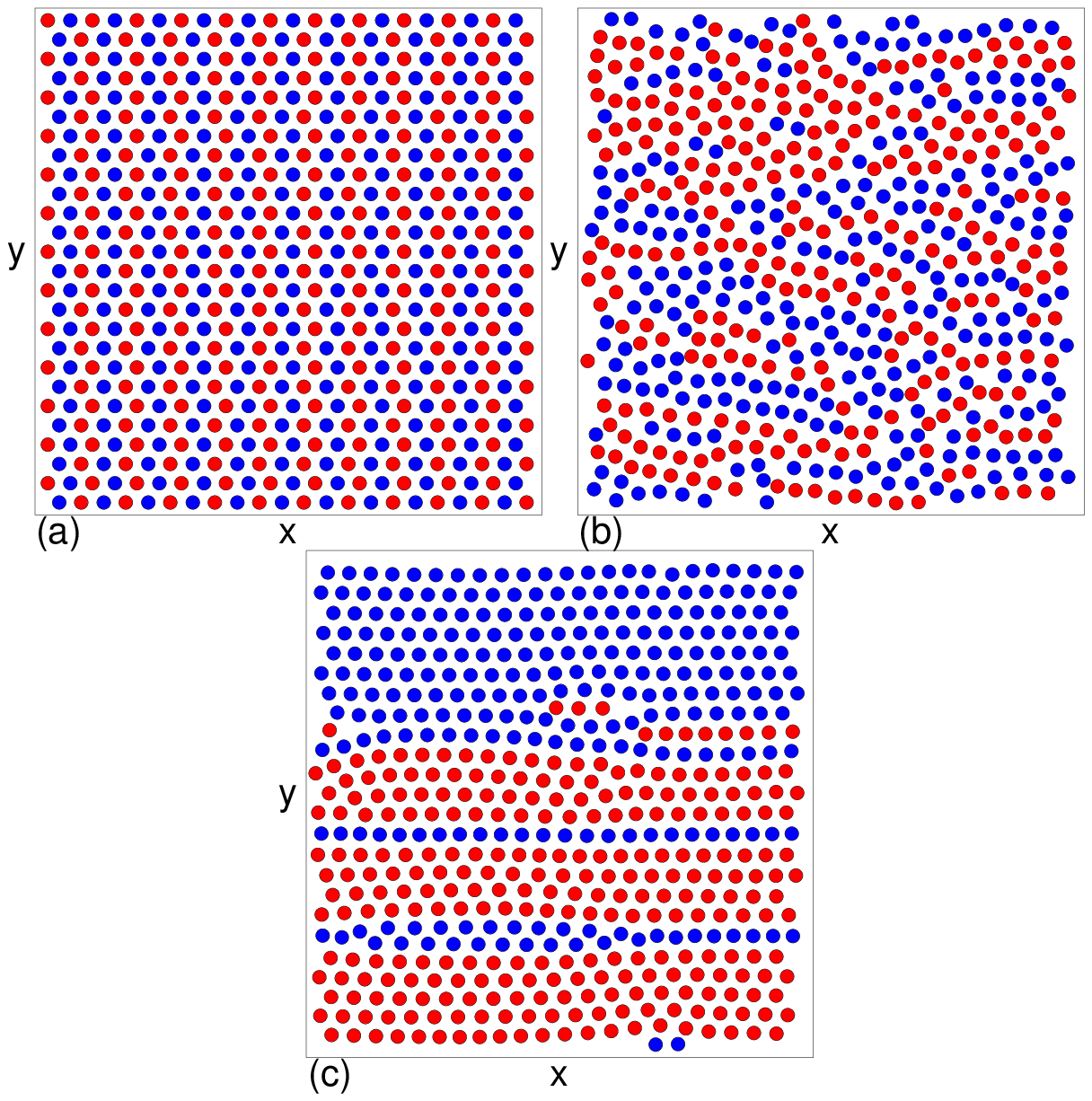}
\caption{Particle positions
for the system from Fig.~\ref{fig:1} at $\rho = 0.441$ and $B = 0.0$,
where the particles have only a Coulomb repulsion.
Red (blue) particles are driven in the
negative (positive) $x$ direction.
(a) The jammed phase J at $F_D = 0.05$.
(b) The moving liquid ML at $F_D = 0.4$.
(c) The segregated laned state L at $F_D = 1.5$.
Note that the radius of the circles representing
the particles has been chosen for visual
clarity here and throughout this work.
}
\label{fig:2}
\end{figure}

We first consider particles with a purely repulsive Coulomb
interaction, obtained by setting $B = 0.0$.
The behavior is expected to be similar to what has been found
in previous studies of repulsive laning particles, such as
for particles
with Yukawa interactions \cite{Dzubiella02a,Chakrabarti04,Glanz12}.
In Fig.~\ref{fig:1}(a), we plot
the velocity-force curve for a system with $\rho = 0.441$.
At low drives there is a jammed region
with $\langle V \rangle = 0.0$,
and the unjamming transition or onset of motion
occurs near $F_D = 0.225$, where $\langle V \rangle$ becomes finite.
A fluctuating velocity regime
of disordered or moving fluid flow
occurs for $0.225 < F_D < 0.78$,
and there is a high-drive linear region for
$F_D > 0.78$ where the velocity fluctuations are strongly reduced and the system
enters a laned state.
Figure~\ref{fig:1}(b) shows the corresponding $d\langle V \rangle/dF_D$
versus $F_D$ curve, where we have performed a running average over
five driving force increments to smooth the curve.
This measure exhibits a clear peak near the jammed-unjammed transition,
has strong fluctuations in the moving fluid regime, and shows a
final cusp near the transition to the laned state.
In Fig.~\ref{fig:1}(c)
we plot the fraction of sixfold coordinated particles
$P_6=N^{-1}\sum_i^N\delta(z_i-6)$ versus $F_D$, where
the particle coordination number $z_i$ is obtained
from a Voronoi construction of the particle positions.
In the jammed phase, $P_6$ is close to 1.0,
consistent with the system forming a jammed triangular lattice,
as shown in Fig.~\ref{fig:2}(a) at $F_D = 0.05$.
At the unjamming transition, $P_6$ drops
when the system breaks up into a partially segregated fluid-like state,
shown in Fig.~\ref{fig:2}(b) at $F_D = 0.4$.
In this moving liquid phase, the triangular ordering is lost,
and there are numerous topological defects along with
large velocity fluctuations.
Additionally, the frequent collisions between particles that occur
in this fluctuating phase cause
$\langle V \rangle/\eta F_D$ to be much less than one.
For a completely free particle interacting with no
other particles, we would have $\langle V \rangle/\eta F_D = 1.0$.
The onset of the laned state correlates with a jump up in $P_6$ and
is also where $d\langle V \rangle/dF_D$ approaches 1.0, which
indicates reduced drag.
Figure~\ref{fig:2}(c) shows the particle configurations
in the laned state at $F_D = 1.5$, where large blocks of particles are all
moving in the same direction along well-defined lanes.
There is triangular ordering of the particles
within the lanes, which leads to the increase in $P_6$ to $P_6=0.9$ at the
onset of the laned state in
Fig.~\ref{fig:1}(c).
There is not complete ordering of the particles,
since the boundaries separating lanes moving in opposite directions
contain aligned dislocations
that glide as the crystallites slip past one another.

\begin{figure}
\includegraphics[width=\columnwidth]{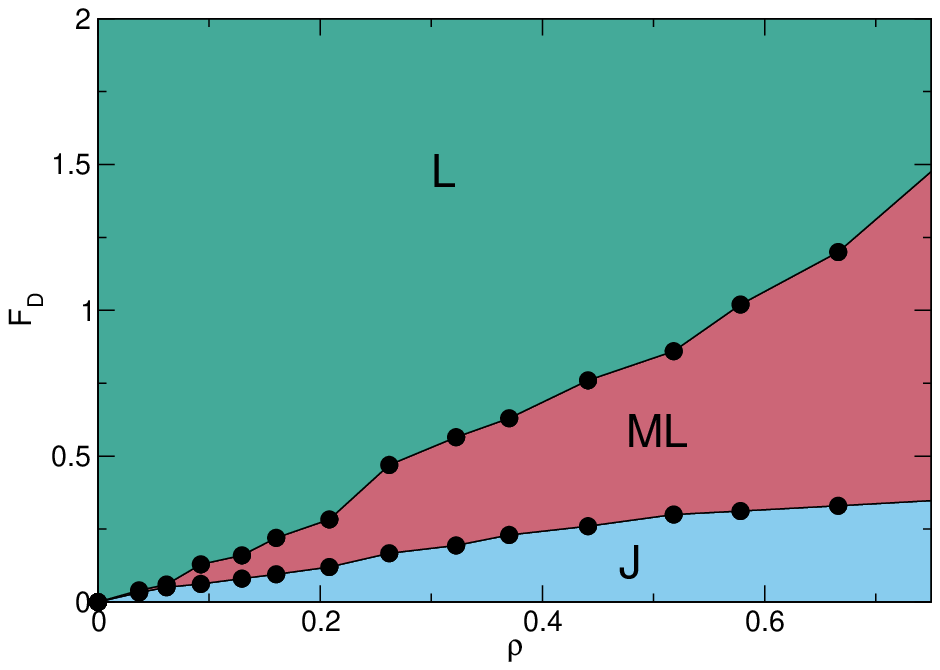}
\caption{Dynamic phase diagram as a function of $F_D$ vs $\rho$
for the system from Fig.~\ref{fig:1} with $B = 0.0$.
J: jammed state (blue);
ML: moving liquid (red);
and L: the higher drive laned state (green).
}
\label{fig:3}
\end{figure}

\begin{figure}
\includegraphics[width=\columnwidth]{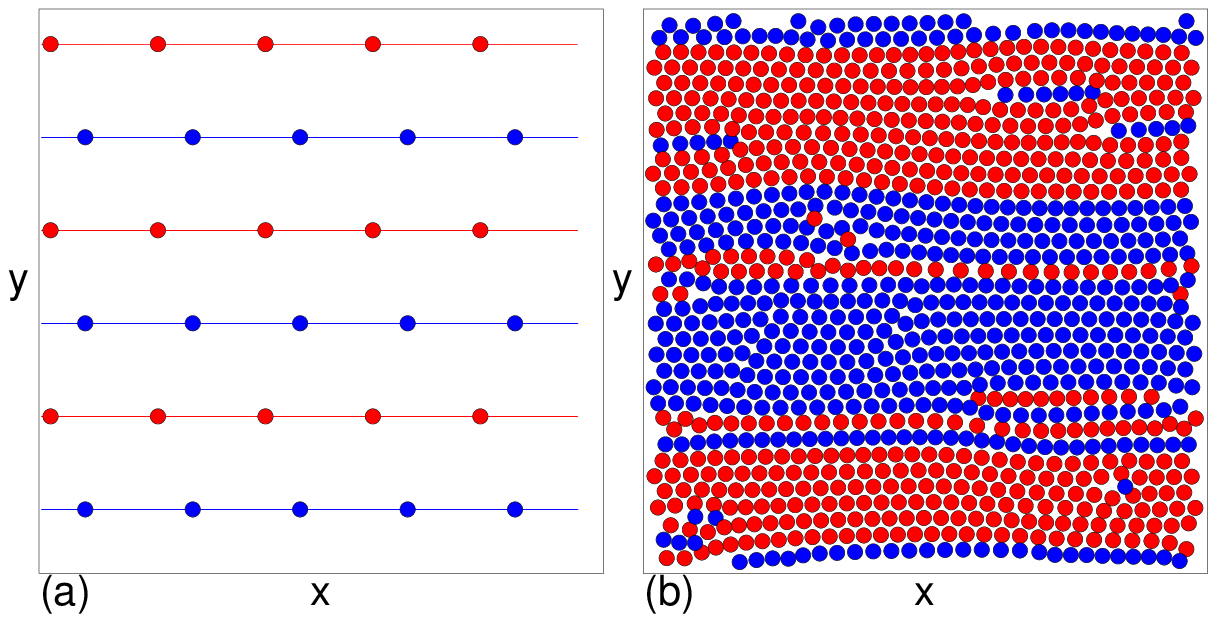}
\caption{Particle positions for the system from Fig.~\ref{fig:3}.
Red (blue) particles are driven in the
negative (positive) $x$ direction.
(a) At $\rho = 0.037$ and $F_D = 0.5$,
the system has fully segregated into one-dimensional (1D) lanes.
Here the particle trajectories are
also plotted as colored lines indicating
the direction of drive, with red for $\sigma_i=-1$ and blue for
$\sigma_i=+1$.
(b) The laned state at
$\rho = 0.86$ and
$F_D = 1.6$.
}
\label{fig:4}
\end{figure}

\begin{figure}
\includegraphics[width=\columnwidth]{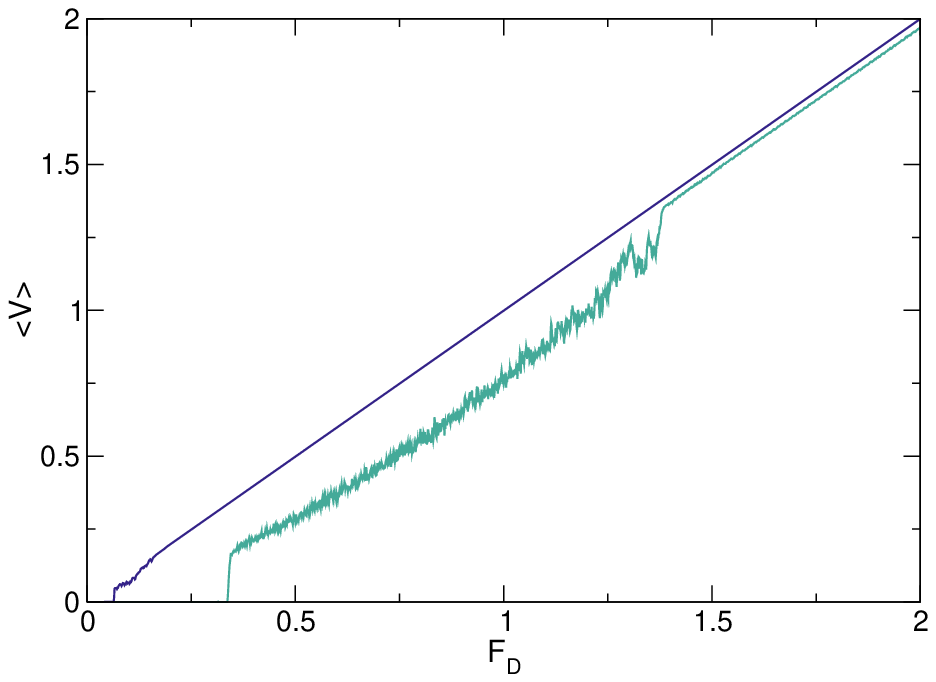} 
\caption{$\langle V \rangle$ vs $F_D$ for the system
from Fig.~\ref{fig:4} with $B=0.0$ at
$\rho = 0.037$ (blue) and $\rho = 0.86$ (green).
For the higher density $\rho = 0.86$,
there is an extended regime of the moving liquid state
and a sharper transition into the laned state at higher drives.
}
\label{fig:5}
\end{figure}

We performed a series of simulations for the system in
Fig.~\ref{fig:1} at varied particle density $\rho$,
and from the features in the transport curves and $P_6$,
we map out the evolution of the three phases, shown
in Fig.~\ref{fig:3} as a function of $F_D$ versus $\rho$.
The critical unjamming force $F_c$ separating the jammed and moving liquid
phases increases
with increasing $\rho$,
and the drive at which the onset of the laning phase occurs
also increases.
At lower $\rho$, we find some states where the system can form
single-file one-dimensional (1D) lanes,
as shown in Fig.~\ref{fig:4}(a) at $\rho = 0.037$ and $F_D = 0.5$.
When the density is higher, the lanes are wider, as illustrated
in Fig.~\ref{fig:4}(b) for $\rho = 1.0$ at $F_D = 1.6$.
The emergence of thinner lanes for lower $\rho$ may reflect the fact that
the smaller number of particles present can dynamically anneal into a well-ordered
state very effectively, and/or that thicker lanes are destabilized by
the Coulomb repulsion.
In Fig.~\ref{fig:5}, we plot $\langle V \rangle$ versus
$F_D$ for the system from Fig.~\ref{fig:4} at $\rho = 0.037$ and $\rho = 0.86$.
For the higher density, there is an extended range of disordered flow
and a sharper transition into the laned state,
as indicated by the jump up in $\langle V \rangle$.

Wide lanes were also observed in systems
of oppositely moving disks \cite{Reichhardt18};
however, one difference is that the jammed phases
for the disks are highly heterogeneous and are composed
of coexisting high- and low-density regions. In contrast,
for the Coulomb system, the jammed state is homogeneous. This is because
the long-range Coulomb
repulsion penalizes significant density fluctuations in the
system.
The disk system also had
a critical density below which jamming does not occur \cite{Reichhardt18},
while for the Coulomb system, a jammed phase is present
down to arbitrarily low densities.
This is also a consequence of the long-range Coulomb interactions.
Even at zero drive in two dimensions (2D),
there is a finite density 
for solidification or jamming transitions to occur in hard disk systems
\cite{Reichhardt14}. 
In work on gliding dislocations moving in opposite directions,
it was found that due to the long-range strain field,
the dislocations could jam at arbitrarily low densities \cite{Dahmen11},
which is consistent with the
idea that it is always possible to reach a jammed state in a
Coulomb system.
We also note that features in the velocity-force curves, peaks in
$d\langle V\rangle/dF_D$, and changes in $P_6$
have been used previously to identify different dynamic phases and
transitions between these phases in
driven particle systems with quenched disorder \cite{Reichhardt17}.

\section{Stripe State}

\begin{figure}
\includegraphics[width=\columnwidth]{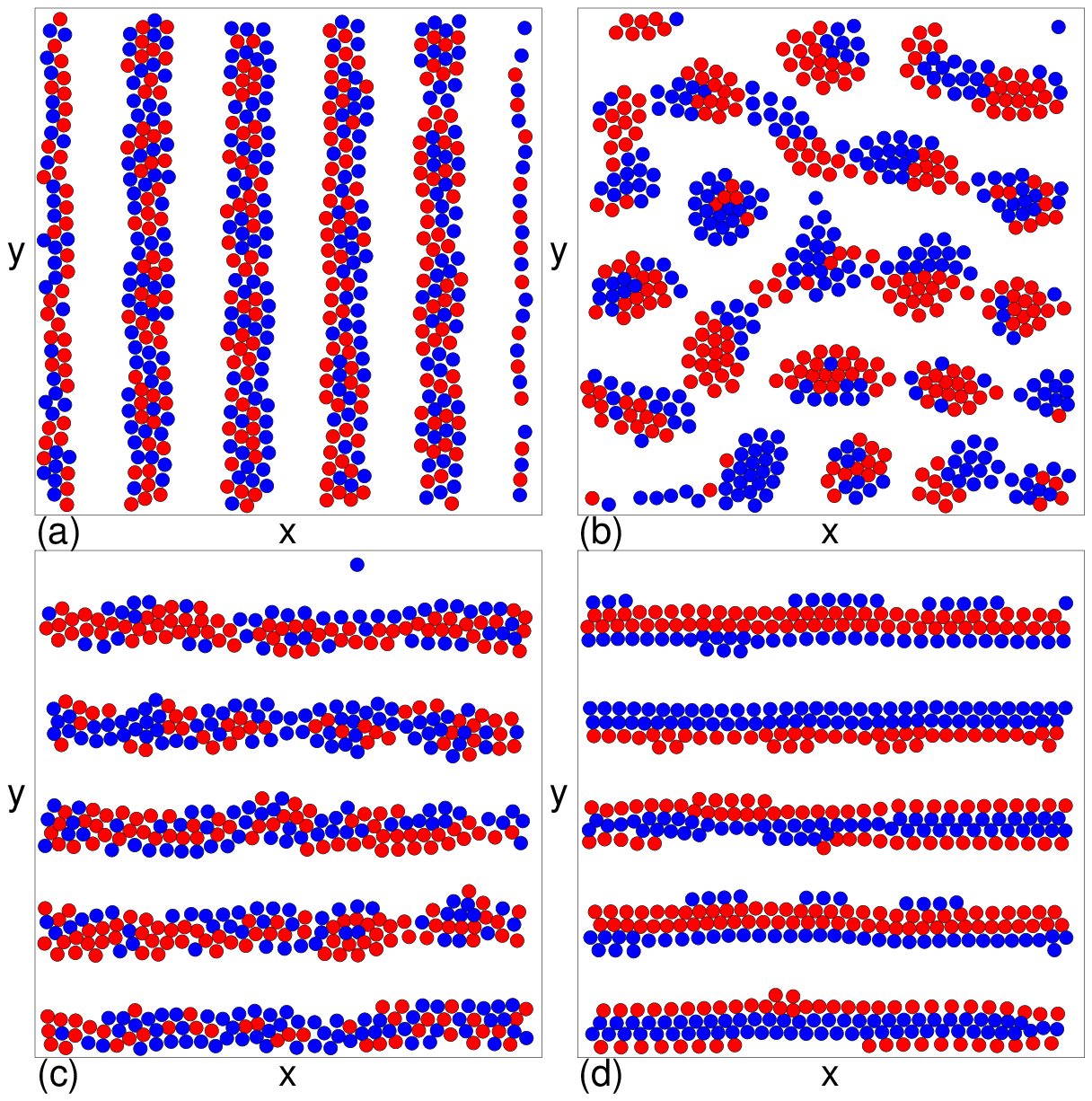}
\caption{Particle positions for the stripe regime at $B=2.1$ and $\rho=0.441$
in different dynamical phases.
Red (blue) particles are driven in the negative (positive) $x$ direction.
(a) The stripe state at $F_{D} = 0.0$.
(b) A disordered state at $F_{D} = 0.072$. 
(c) The disordered stripe phase at $F_{D} = 0.6$.
(d) A laned stripe state at $F_{D} = 1.0$.
	}
\label{fig:6}
\end{figure}

\begin{figure}
\includegraphics[width=\columnwidth]{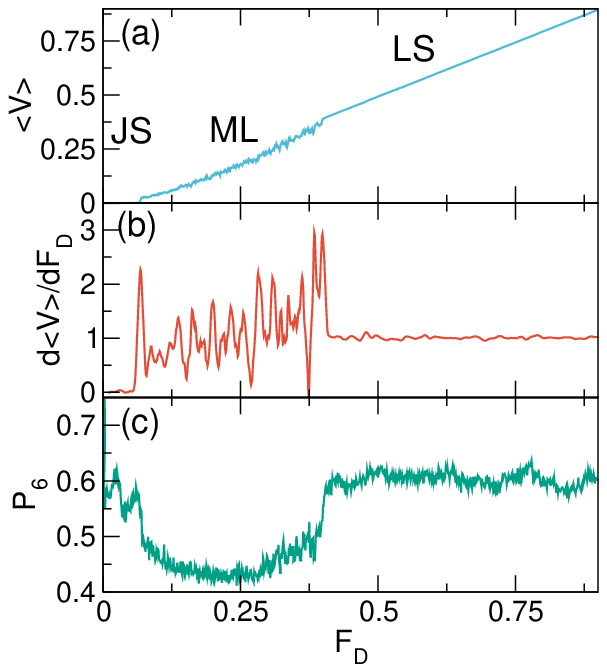}
\caption{$\langle V \rangle$ vs $F_D$ curves for the
stripe-forming system from Fig.~\ref{fig:6} with $B=2.1$ and $\rho=0.441$.
(b) The corresponding $d\langle V \rangle/dF_D$ vs $F_D$.
(c) The corresponding $P_6$ vs $F_D$.
}
\label{fig:7}
\end{figure}

We next consider particles that have both repulsive and attractive
interactions.
For $B<1.9$, the dynamic phases we observe are similar to those described
in Sec.~III for the $B=0.0$ system, with
a jammed uniform state, a moving liquid state, and a higher-drive laned state.
For $1.9 < B < 2.25$, the system forms a jammed stripe state,
as shown in Fig.~\ref{fig:6}(a) at $F_D = 0.0$ for $B = 2.1$ and
$\rho = 0.441$.
As $F_D$ is increased, the jammed phase breaks up into a
disordered phase, illustrated in Fig.~\ref{fig:6}(b) at $F_D = 0.072$.
At higher drives, stripes reform, but these stripes are are now oriented
in the direction of the drive, as shown in Fig.~\ref{fig:6}(c) at
$F_D = 0.6$.
The oriented stripes fluctuate rapidly and contain a varying number of
particles that are exchanging places, so that the system is in
a fluctuating quasi-one-dimensional fluid rather than a laned state.
At the highest drives, the stripes settle into well-ordered stable
lanes where the particles within each stripe travel in opposite directions
to each other along segregated lanes,
as imaged in Fig.~\ref{fig:6}(d) at $F_D = 1.0$.
The transport curve signatures produced by the different phases are shown
in Fig.~\ref{fig:7}(a), where we plot
$\langle V\rangle$ versus $F_D$.
The transitions are more clearly highlighted
in the corresponding plot of
$d\langle V \rangle/dF_D$ versus $F_D$ in Fig.~\ref{fig:7}(b),
where multiple peaks appear.
In the laned stripe phase,
$d\langle V \rangle/dF_D$ is close to one, indicating almost linear flow.
The amount of topological order varies somewhat in the different dynamic
stripe states, which are visible in the $P_6$ versus $F_D$ curve
shown in Fig.~\ref{fig:7}(c). The order is lowest in the disordered
flow state, but even in the laned stripe phase, the value of $P_6$ is lower
than in the corresponding laned phase for the $B=0.0$ system in
Fig.~\ref{fig:1}(c).
The laned stripe phase appears when $1.9 < B < 2.35$,
and also occurs for values of $B$
at which the jammed state is a bubble state rather than a stripe
state. This occurs because the application of driving
can stretch the bubbles into stripes.
From the transport curves and $P_6$, it is not possible to distinguish the
2D disordered fluctuating state just above unjamming from
the disordered stripe state, so we lump the 2D moving liquid and
quasi-one-dimensional moving liquid together into a
disordered moving liquid phase. We note that it may be possible to use
machine learning techniques to delineate a transition between these two
liquid states \cite{Reichhardt25}.

\begin{figure}
\includegraphics[width=\columnwidth]{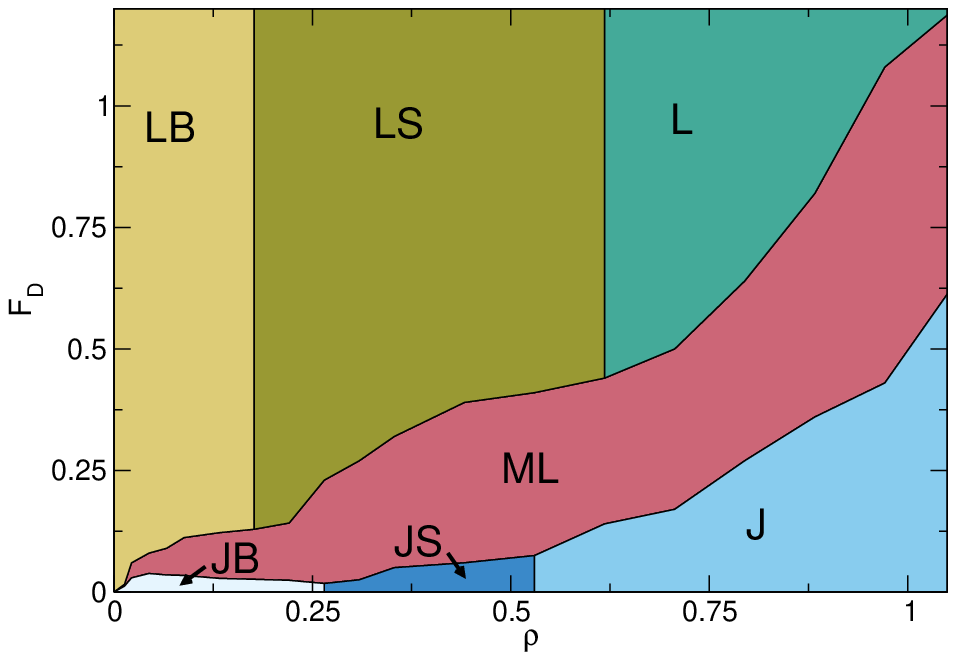}
\caption{Dynamic phase diagram as a function of $F_{D}$ vs $\rho$
for a system with $B = 2.0$.
JB: jammed bubble phase (pale blue),
JS: jammed stripe phase (dark blue),
J: jammed uniform phase (medium blue),
ML: disordered moving liquid phase (red),
LB: laned bubble phase (yellow),
LS: laned stripe phase (olive green),
and L: laned state (green).
}
\label{fig:8}
\end{figure}

\begin{figure}
\includegraphics[width=\columnwidth]{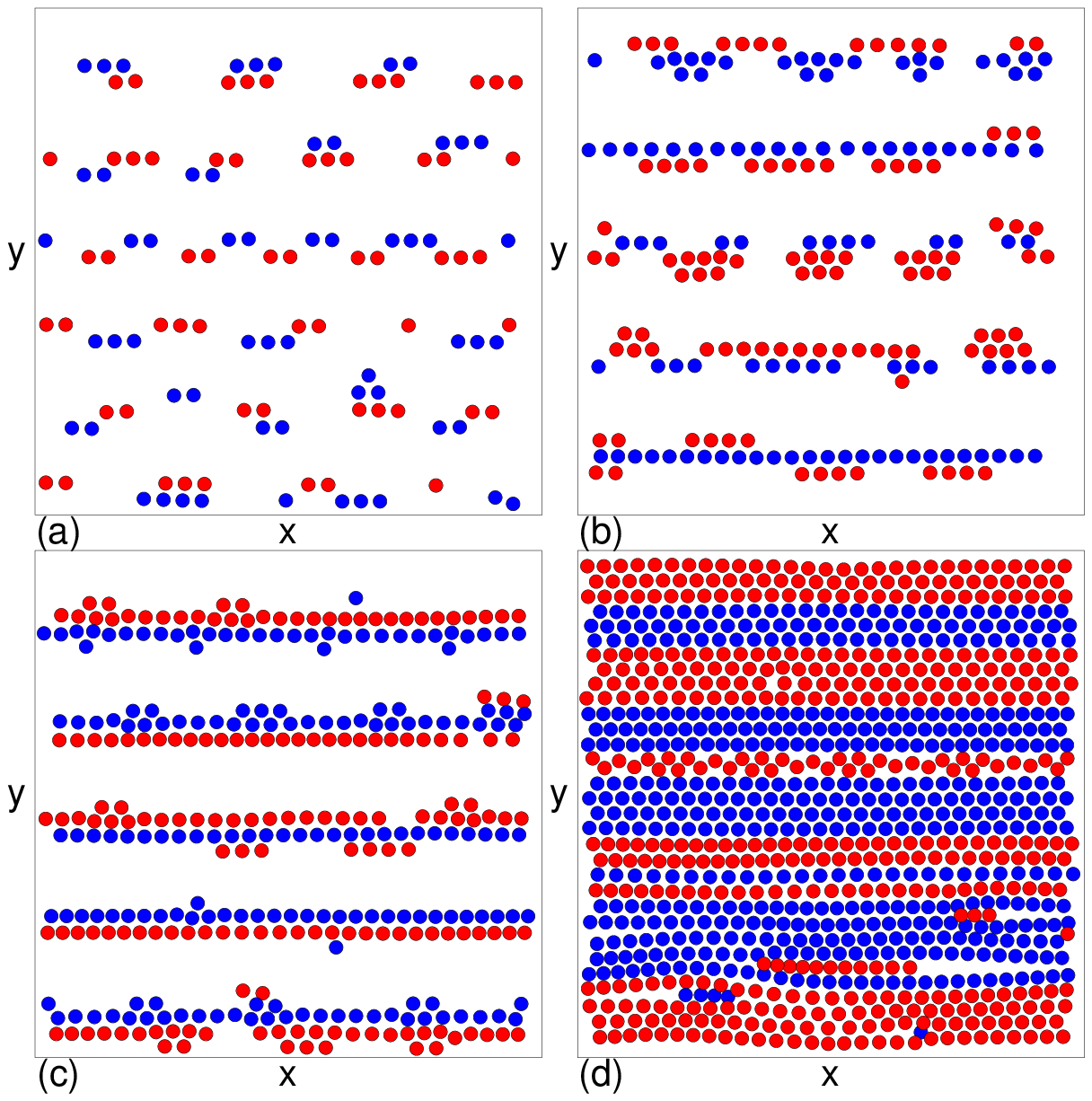}
\caption{Particle positions for the $B=2.0$ system from Fig.~\ref{fig:8}.
Red (blue) particles are driven in the negative (positive) $x$ direction.  
(a) The laned bubble state at $\rho = 0.0926$ and $F_D=0.25$.
(b) The laned bubble state at $\rho = 0.16$ and $F_D=0.25$,
where partial stripe ordering appears.
(c) The laned stripe state at $\rho = 0.26$ and $F_D=0.5$.
(d) The uniform laned state at $\rho = 0.76$ and $F_D=1.0$.
}     
\label{fig:9}
\end{figure}

In Fig.~\ref{fig:8}, we map out the dynamic phase diagram
as a function of $F_D$ versus $\rho$ for a system with $B = 2.0$.
At finite $B$, the system undergoes
transitions between different ground states
as a function of $\rho$ when $F_D = 0.0$.
When $B = 2.0$,
the system shows a jammed bubble state for $\rho < 0.25$,
a stripe phase for $0.25 \leq \rho \leq 0.53$,
and a uniform jammed phase for $\rho > 0.53$.
The unjamming transition is nonmonotonic as a function of $\rho$
in Fig.~\ref{fig:8}, initially rising in the bubble state,
passing through a local minimum near the
jammed bubble to jammed stripe transition,
and then increasing more rapidly in the uniform jammed state.
The jammed bubble phase unjams into a disordered state,
and at higher drives, a laned bubble state arises,
as illustrated in Fig.~\ref{fig:9}(a) at $\rho = 0.0926$ and $F_D=0.25$.
For the same drive of $F_D=0.25$ at $\rho=0.16$, Fig.~\ref{fig:9}(b)
shows that 
there is some partial stripe ordering.
An ordered stripe state appears at $F_D=0.25$ and $\rho=0.26$ in
Fig.~\ref{fig:9}(c), while
Fig.~\ref{fig:9}(d) shows a high drive uniform laned state at
$\rho = 0.76$ and $F_D=1.0$.
For $0.17 \leq \rho \leq 0.65$, the high-drive state is a laned stripe state
instead of a laned uniform state.
Due to the fact that the drive elongates the bubbles into stripes,
the extent of the laned stripe state exceeds that of the
$F_D=0$ stripe state, which spans only the range
$0.25 \leq \rho \leq 0.53$.

\section{Varied Interactions}

\begin{figure}
\includegraphics[width=\columnwidth]{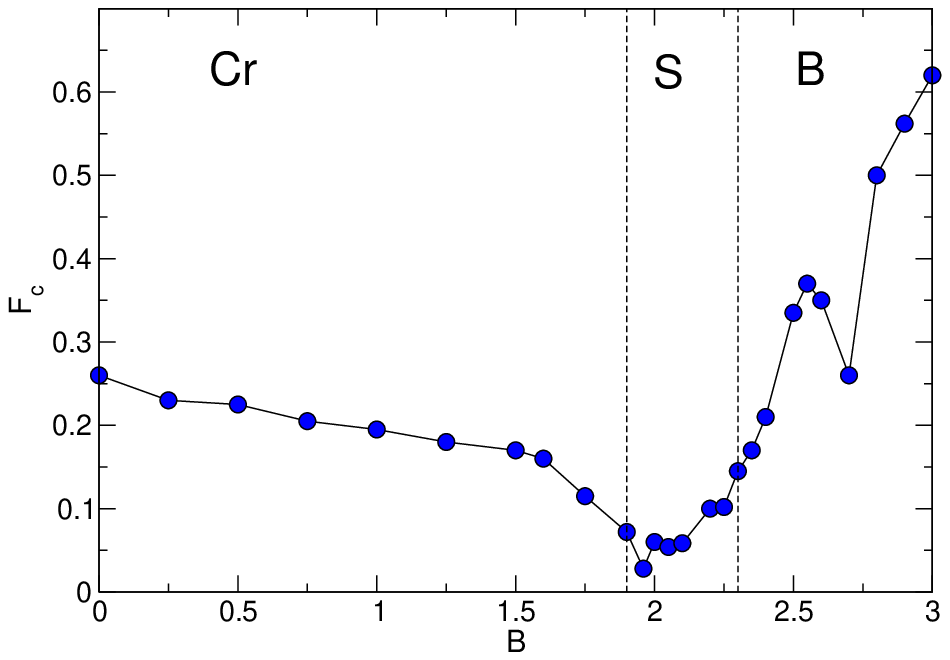}
\caption{The critical unjamming force $F_{c}$ vs $B$
for a system with $\rho = 0.441$. 
The vertical dashed lines indicate where
the system forms jammed crystal (Cr), stripe (S), and
bubble (B) states. $F_c$ varies non-monotonically with $B$.
}
\label{fig:10}
\end{figure}

We next consider systems with fixed $\rho = 0.44$ and
varied $B$ ranging from the purely repulsive $B=0.0$ case to
large $B$ values for which a bubble state is present.
In the absence of a drive,
the system forms a uniform crystal for $0 < B \leq 1.9$,
stripes for $1.9 \leq B < 2.3$,
and bubbles for $B \geq 2.3$.
In Fig.~\ref{fig:10},
we plot the critical unjamming force $F_c$ versus $\rho$, with
vertical lines highlighting the
locations of the $F_D=0$ crystal, stripe, and bubble phases.
The unjamming force $F_c$ is a strongly nonmonotonic
function of $B$ and is affected by the type of pattern that forms
in the system.
There is a gradual decrease of $F_c$ with increasing $B$ throughout
the uniform crystal phase, followed by a more rapid drop to a local
minimum in $F_c$ on the low-$B$ end of the stripe state
near $B = 2.0$.
As $B$ continues to increase,
$F_c$ grows throughout the stripe phase
and then begins to increases more rapidly once the system transitions
into the bubble phase.
A local dip in $F_c$ appears in the bubble phase near $B = 2.7$
corresponding to the onset of plastic bubble depinning,
where particles remain trapped in individual bubbles but the bubbles
themselves move past each other.
In previous work, a similar nonmonotonic depinning threshold
was observed for SALR particles interacting
with a periodic one-dimensional substrate, where
the depinning force reaches a maximum in the stripe phase and then
falls off rapidly in the 
bubble phase \cite{Reichhardt24}.
This is consistent with what we find in Fig.~\ref{fig:10},
where the stripe phase is the softest phase
and has the lowest elastic constant, which causes the
unjamming force to be low
but would also produce a local maximum in the depinning force
when a substrate is present.
The phase we observe here is a jammed rather than a pinned state,
since the system contains no quenched disorder.
The jamming arises purely from the interaction of the particles that are
attempting to move in opposite directions.
It is possible that for values of $B$ larger than those considered here,
$F_c$ would drop again when the radii of the bubbles begins to shrink due
to the very strong attractive term.

\begin{figure}
\includegraphics[width=\columnwidth]{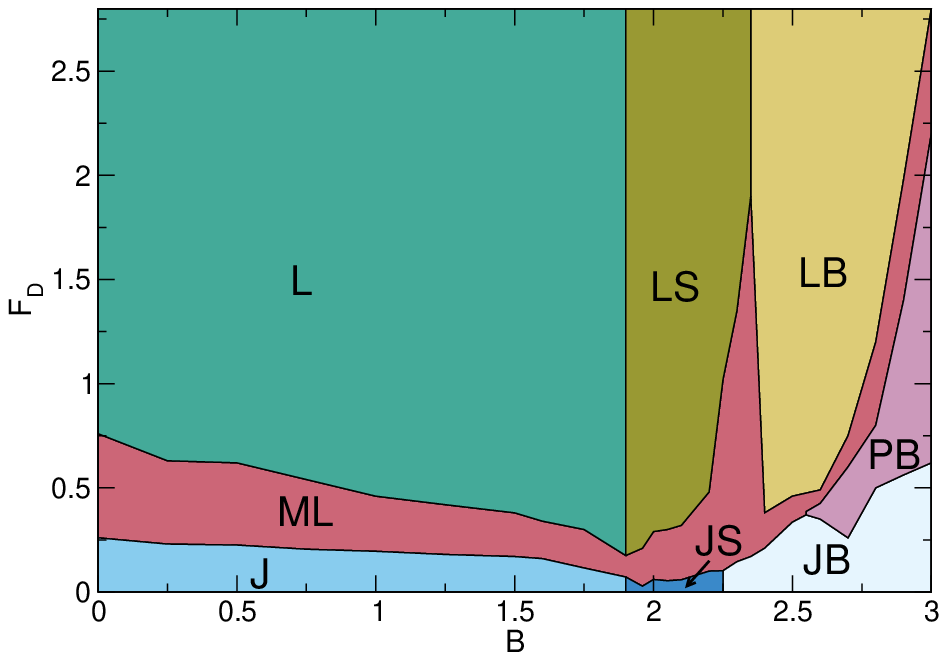}
\caption{Dynamic phase diagram as a function of $F_D$ vs $B$
for a system with $\rho = 0.441$.
JB: jammed bubble phase (pale blue),
JS: jammed stripe phase (dark blue),
J: jammed uniform phase (medium blue),
ML: disordered moving liquid phase (red),
LB: laned bubble phase (yellow),
LS: laned stripe phase (olive green),
L: laned state (green),
and PB: plastic bubble phase (purple).
}
\label{fig:11}
\end{figure}

\begin{figure}
\includegraphics[width=\columnwidth]{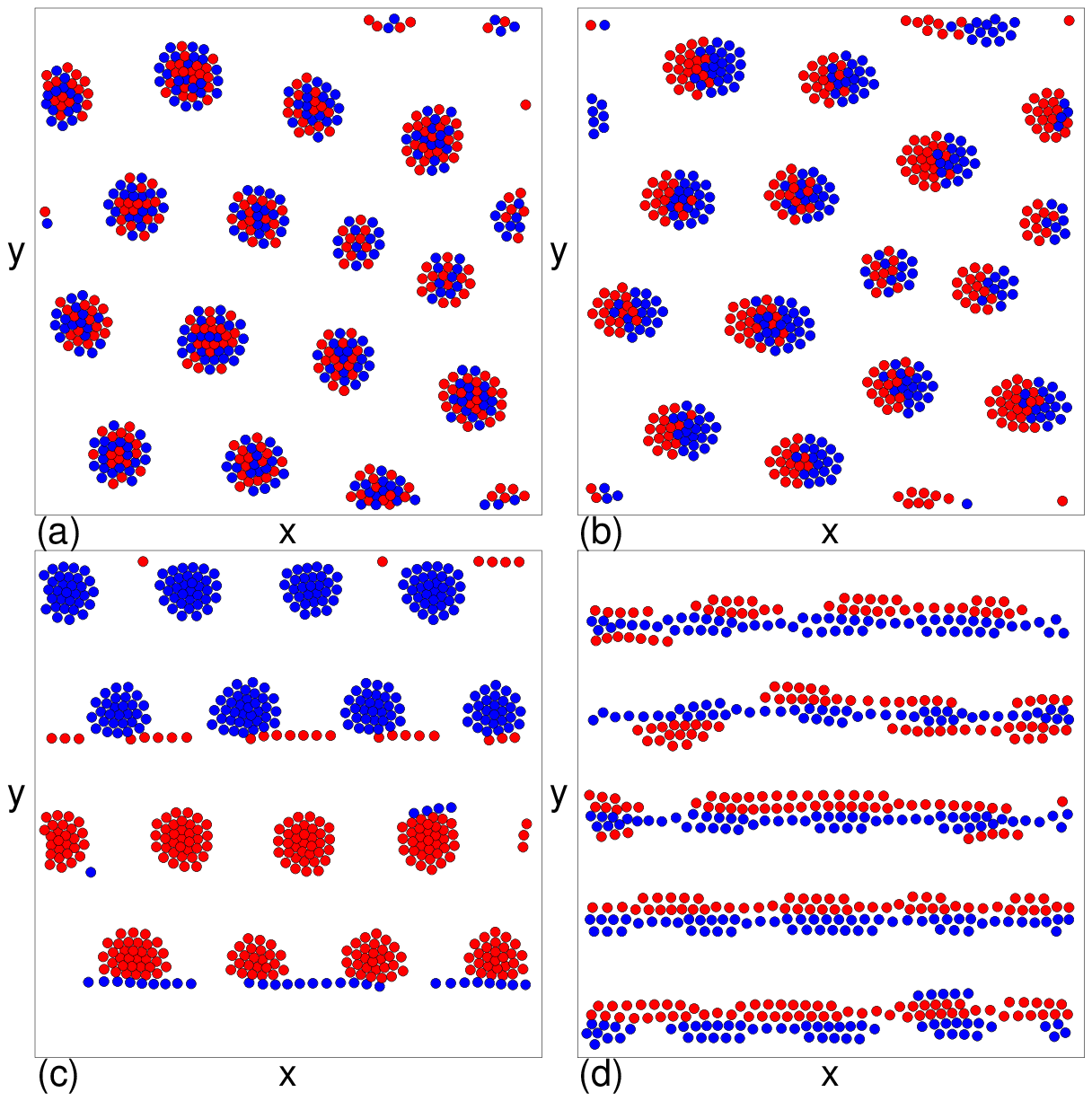}
\caption{Particle positions for the system in Fig.~\ref{fig:11}
with $\rho=0.441$.
Red (blue) particles are driven in the negative (positive) $x$
direction.
(a) Spatially mixed bubbles at $F_D = 0.03$ and $B = 2.35$.
(b) Polarized or segregated bubbles at $F_D = 0.1$ and $B = 2.35$.
(c) The laned bubble state with a small number of stripes at
$F_D = 2.5$ and $B=2.35$.
(d) The laned state at $F_D=1.5$ and a smaller $B = 2.25$.
}
\label{fig:12}
\end{figure}

\begin{figure}
\includegraphics[width=\columnwidth]{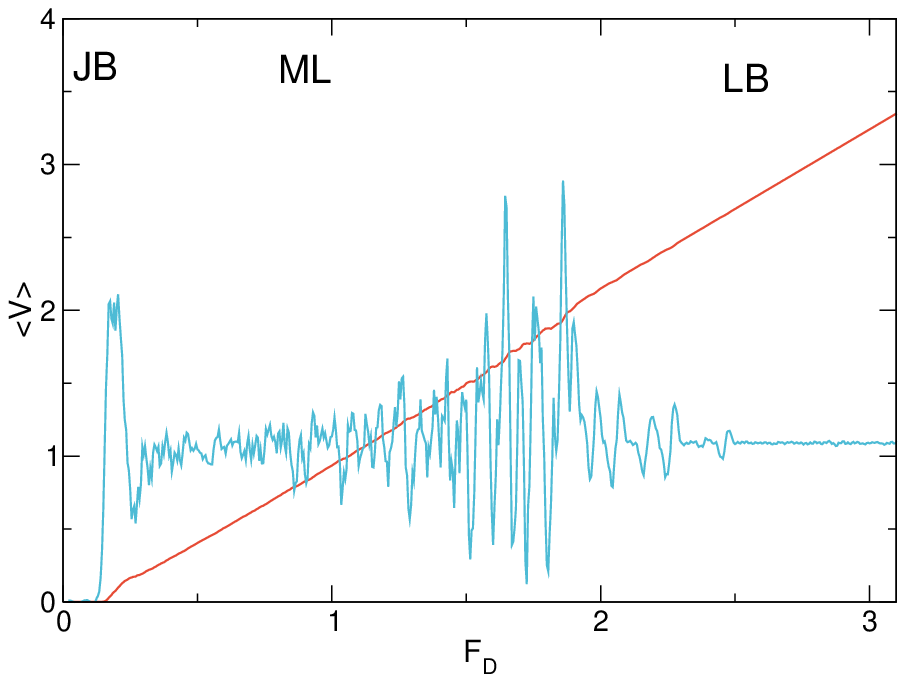}
\caption{$\langle V \rangle$ (red)
and $d\langle V \rangle/dF_{D}$  (blue) vs $F_D$
for the system from Figs.~\ref{fig:11} and \ref{fig:12}
with $\rho=0.441$
at $B = 2.35$,
showing the jammed bubble (JB),
fluctuating moving liquid (ML), and laned bubble (LB) states.
}
\label{fig:13}
\end{figure}

Based on the transport curves, $P_{6}$, and images of the particles,
we construct a dynamic phase diagram as a function of
$F_{D}$ versus $B$ for the $\rho = 0.441$ system in
Fig.~\ref{fig:11}.
For $B < 1.9$, there is a uniform jammed phase, a disordered
moving liquid phase, and a uniform laned phase, similar to
what is found for the $B=0.0$ system 
in Fig.~\ref{fig:2}.
For $1.9 < B < 2.3$, there is a jammed stripe phase,
a disordered moving liquid phase, and
an ordered laned stripe state.
The laned stripe phase persists up to $B = 2.3$,
even though a jammed bubble state forms for $B > 2.2$.
This occurs because the bubbles become elongated
in the driven state and merge to form a stripe.
The drive at which
the ordered laned stripe phase
arises passes through a peak at $B = 2.35$,
which also corresponds to the value of $B$ that separates
the lower $B$ laned stripe state from the
higher $B$ laned bubble state.
At $B = 2.35$, the jammed bubble state contains bubbles in which the
two particle species are evenly spatially distributed,
as shown in Fig.~\ref{fig:12}(a) for $F_{D} = 0.03$. 
As the drive increases,
the particles gradually rearrange 
to form a polarized jammed bubble state, with the $\sigma=-1$ particles
accumulating on the left side of each bubble and the $\sigma=+1$ particles
accumulating on the right side of each bubble,
as shown in Fig. 12(b) at $F_{D} = 0.1$. The attractive
interaction term prevents the
bubbles from tearing apart.
At higher drives, the system forms a disordered bubble flow phase in which
the bubbles move past each other,
and eventually a high-drive bubble phase containing a small number of stripes
emerges, as illustrated in Fig.~\ref{fig:12}(c)
at $F_{D} = 2.5$. Here the bubbles have become fully segregated by species.
Figure~\ref{fig:12}(d) shows 
a laned segregated stripe state at $B = 2.25$ and $F_{D} = 1.5$.
The drive at which full species segregation occurs
increases near the boundary between the laned stripe and laned bubble states.
In Fig.~\ref{fig:13}, we plot
$\langle V \rangle$ and $d\langle V \rangle/dF_{D}$ versus $F_D$
for the same system at $B = 2.35$, showing
the jammed bubble, fluctuating moving liquid, and laned bubble states.
For clarity, we have performed a running average
over the nearest five points on the differential mobility curve.

\begin{figure}
\includegraphics[width=\columnwidth]{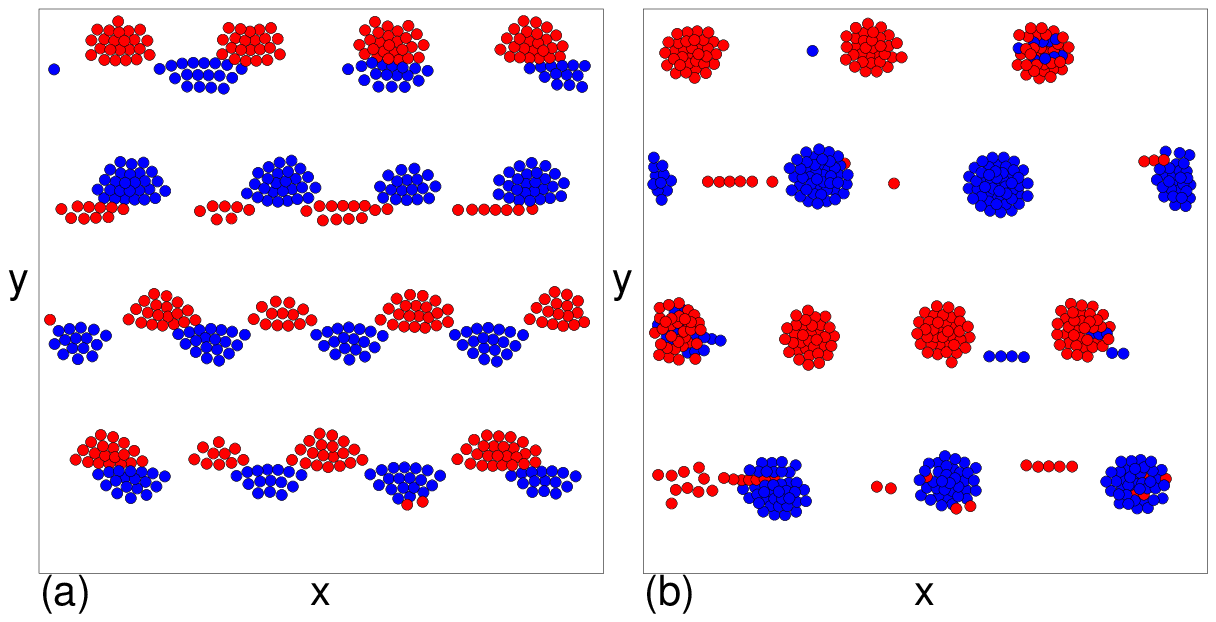}
\caption{Particle positions for the system from Fig.~\ref{fig:11}
with $\rho=0.441$ at $F_D=1.0$ in the laned bubble state.
Red (blue) particles are driven in the negative (positive) $x$
direction.
(a) $B = 2.4$. (b) $B = 2.6$.
}
\label{fig:14}
\end{figure}

For $2.25 < B < 2.6$, the jammed bubble state
unjams into a disordered moving liquid phase,
while for $B \geq 2.6$, the unjamming occurs
via the formation of a plastic bubble phase,
where particles remain trapped in individual bubbles but the bubbles
move past each other.
The segregation of the particles into distinct bubbles can
occur in two ways. In the first, the bubbles tear apart along the direction
perpendicular to the drive, as 
shown in Fig.~\ref{fig:14}(a) for $B = 2.4$ at $F_D = 1.0$.
In the second, the bubbles retain their circular shape and are each largely
composed of a single particle species with a few particles of the
opposite species remaining trapped in each bubble,
as shown in Fig.~\ref{fig:14}(b) for $B = 2.6$ at $F_D = 1.0$.

\begin{figure}
\includegraphics[width=\columnwidth]{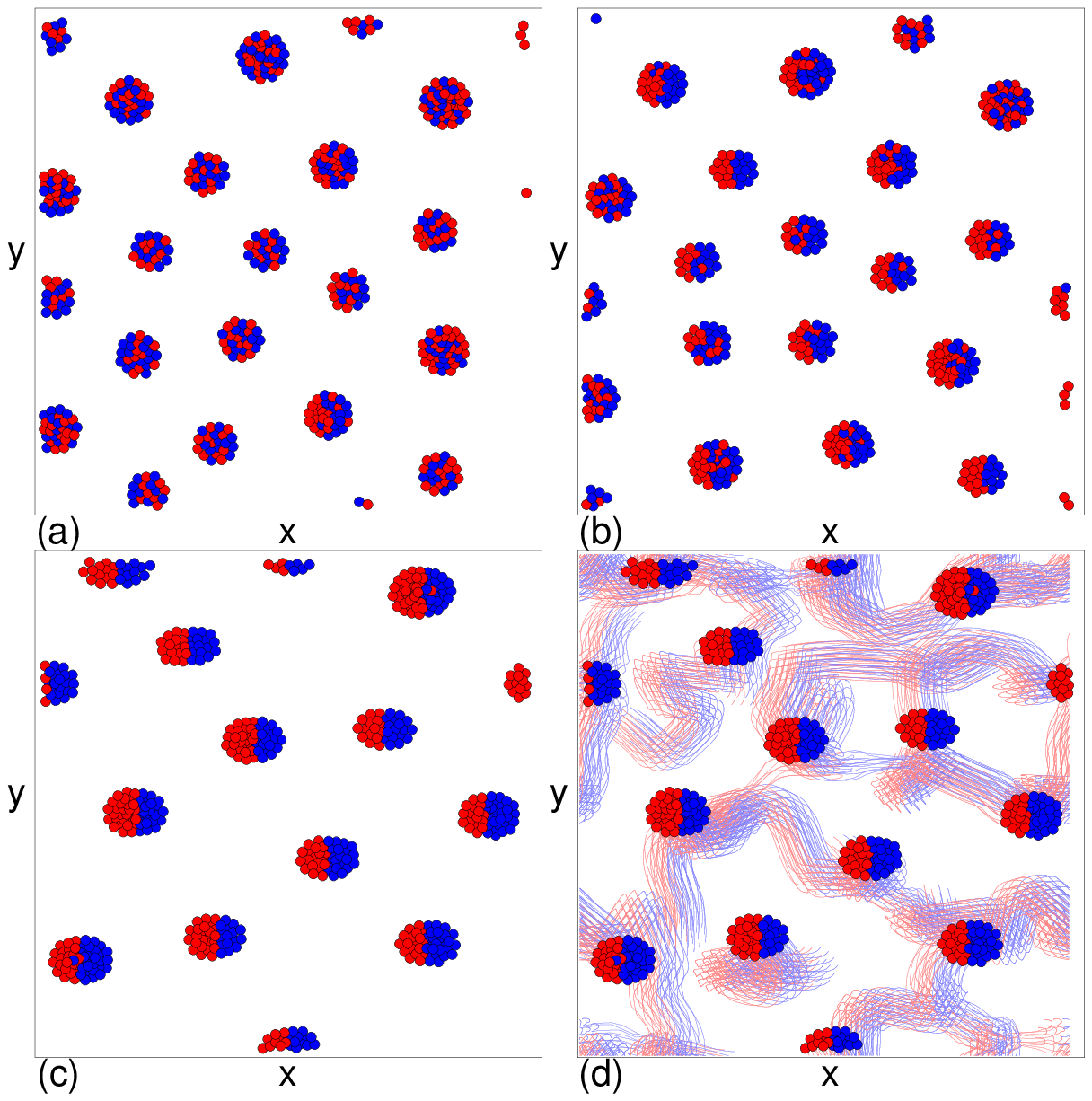}
\caption{Particle positions for the system from Fig.~\ref{fig:11} with
$\rho=0.441$ at $B=2.8$.
Red (blue) particles are driven in the negative (positive) $x$ direction.  
(a) A mixed bubble state at $F_D = 0.0$.
(b) At $F_D = 0.25$, the particles in the bubbles start to rearrange
into a polarized jammed bubble state.
(c) The fully polarized bubble state at $F_D = 0.45$, just below
the unjamming transition.
(d) At $F_D=0.5$ in the plastic bubble phase,
the particles remain confined to the bubbles,
but the bubbles move past one another. Lines illustrate the particle trajectories,
with light red (light blue) lines indicating the motion of particles driven
in the negative (positive) $x$ direction.
}
\label{fig:15} 
\end{figure}

In Fig.~\ref{fig:15}, we show the jammed bubble phase
from the system in Fig.~\ref{fig:11} at $B = 2.8$ and $F_D = 0.0$,
where a bubble lattice appears that is composed of bubbles with spatially
mixed particle species.
As $F_D$ increases, the bubbles undergo rearrangements and the particle
species begin to segregate,
as shown in Fig.~\ref{fig:15}(b) at $F_D = 0.25$, but the system
remains jammed. 
The bubbles become increasingly segregated and elongated
as the drive increases further,
as illustrated in Fig.~\ref{fig:15}(c) at $F_D = 0.45$.
Above the critical unjamming force $F_c$,
the particles remain in the bubbles but the
bubbles begin to move past one another,
as shown in Fig.~\ref{fig:15}(d) at $F_D = 0.5$, where we highlight the
particle locations and trajectories over time.
When $B > 2.6$, as is the case in Fig.~\ref{fig:15},
the particles are so tightly bound to the bubbles that they are unable
to escape from their original bubble.
Since there are different numbers of particles of each species in each
bubble, there is a force imbalance such that
some bubbles experience a net negative $x$ direction force while others
have a net positive $x$ direction force.
This causes the bubbles to move past one another. The long range Coulomb
repulsion between adjacent bubbles is also unbalanced because not all of
the bubbles contain the same total number of particles, and this results
in a distortion of what would otherwise be a roughly triangular lattice
of bubbles.
For higher drives, the system transitions into a disordered moving
liquid state and then reorganizes
into a laned bubble state, as shown in Fig.~\ref{fig:14}(b).

\begin{figure}
\includegraphics[width=\columnwidth]{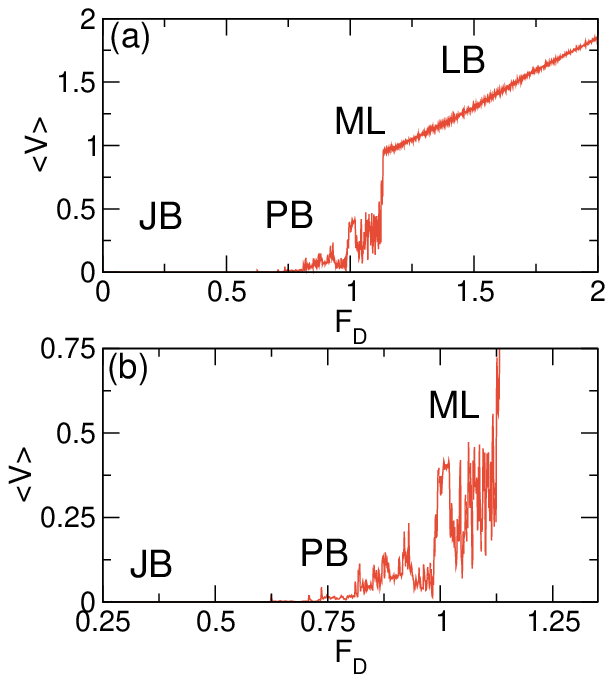}
\caption{(a) $\langle V \rangle$ vs $F_D$ for the system in
Fig.~\ref{fig:15} at $\rho = 0.411$ and $B = 2.8$,
with letters indicating the jammed bubble (JB), plastic bubble (PB),
moving liquid (ML), and laned bubble (LB) phases.
(b) A blowup of the lower drive region from panel (a)
showing the plastic bubble depinning regime in more detail.
}
\label{fig:16}
\end{figure}

In Fig.~\ref{fig:16}(a), we plot
the velocity-force curve $\langle V\rangle$ versus $F_D$
for the system in Fig.~\ref{fig:15}
with $\rho=0.411$ and $B=2.8$ where there is a plastic bubble phase.
Figure~\ref{fig:16}(b) shows a blowup of the
lower drive regime where the plastic bubble motion occurs.
Here, $\langle V \rangle$ remains low since the bubbles
are only slowly moving past each other,
but there are occasional fluctuations in which
sudden bursts of motion occur.
At higher drives, the particles begin
to escape from the bubbles, and the velocity increases as the
system enters the moving liquid state.

\section{Varied Bubble Density}

\begin{figure}
\includegraphics[width=\columnwidth]{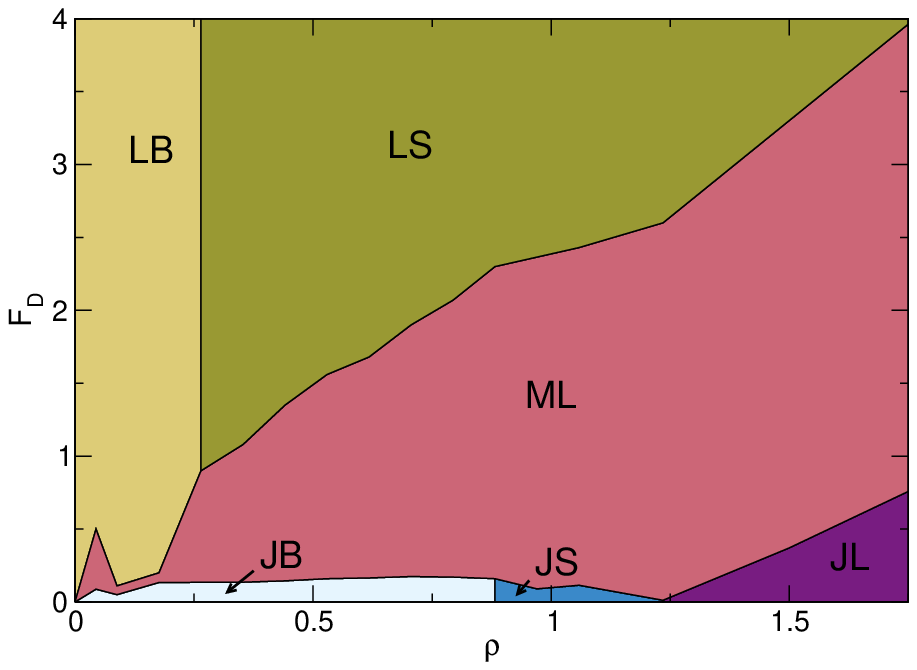}
\caption{Dynamic phase diagram as a function of
$F_D$ vs $\rho$ for a system with $B = 2.3$.
JB: jammed bubble phase (pale blue),
JS: jammed stripe phase (dark blue),
JL: jammed labyrinth phase (dark purple),
ML: disordered moving liquid phase (red),
LB: laned bubble phase (yellow),
and LS: laned stripe phase (olive green).
}
\label{fig:17}
\end{figure}

\begin{figure}
\includegraphics[width=\columnwidth]{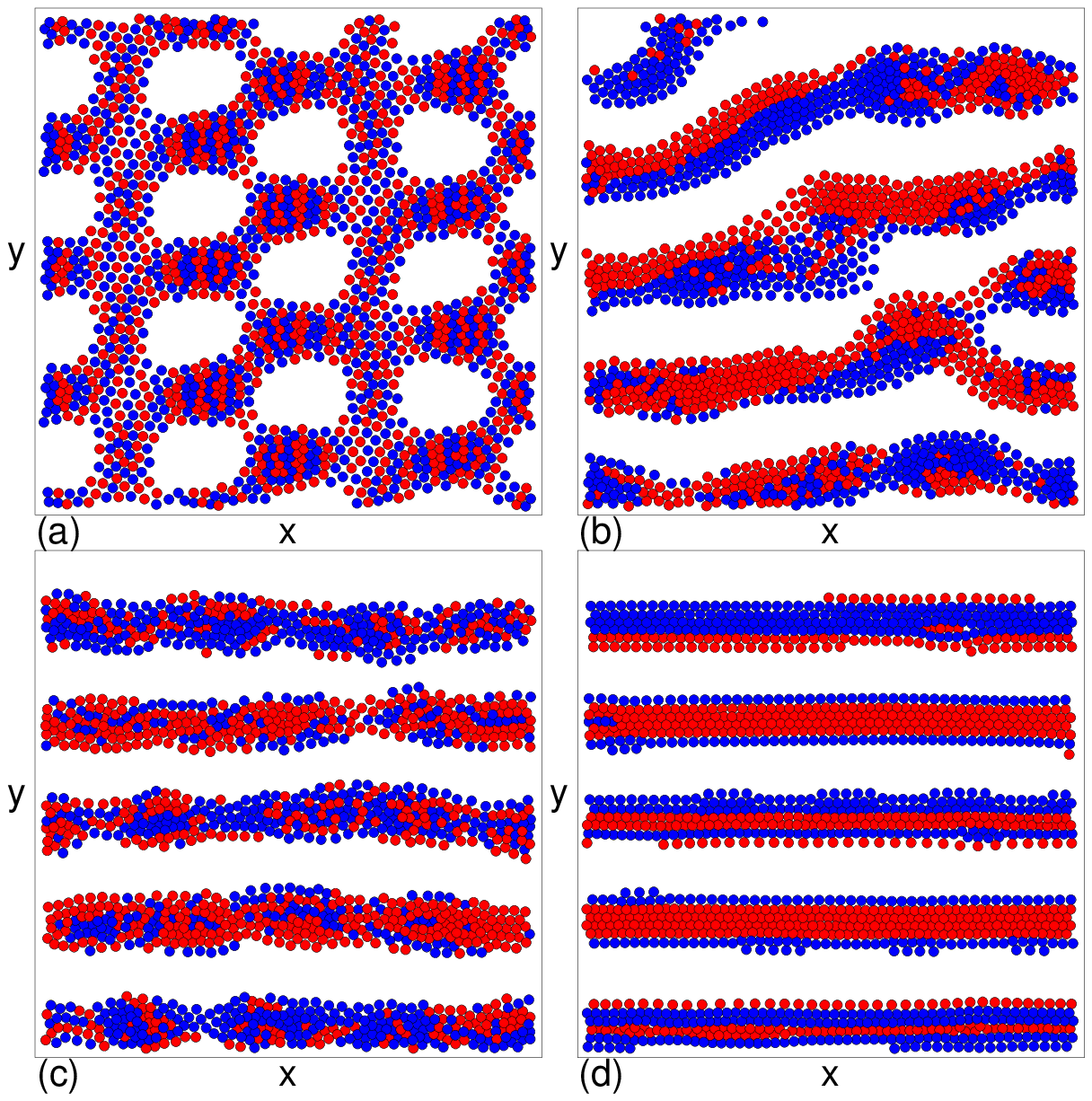}
\caption{Particle positions for the system from Fig.~\ref{fig:17} with
$B=2.3$ at  
$\rho=1.234$.
Red (blue) particles are driven in the negative (positive) $x$ direction.
(a) Jammed labyrinth phase at $F_D = 0.0$.
(b) Disordered moving liquid phase at $F_D = 0.2$.
(c) Disordered moving liquid state at $F_D = 0.8$, where the particles
align with the driving direction. (d) Laned stripe state at $F_D = 3.0$.
}
\label{fig:18}
\end{figure}

We next examine the evolution of the phases for varied density at
$B = 2.3$, where a jammed bubble state appears when $\rho = 0.441$.
In Fig. 17, we show the dynamic phase diagram
as a function of $F_D$ versus $\rho$.
At $F_D=0$, there is a jammed bubble phase for $0 < \rho \leq 0.9$,
a jammed stripe phase for $0.9 < \rho \leq 1.2$,
and a jammed labyrinth for $\rho > 1.2$.
The critical unjamming force is non-monotonic and passes through
a local minimum near the transition from a jammed stripe to
a jammed labyrinth state.
Above the unjamming transition,
there is a disordered moving liquid phase that is
followed at higher drives by a moving laned bubble state for $\rho < 0.25$ and
by a laned stripe phase for $\rho > 0.25$.
The moving liquid phase grows in extent with increasing $\rho$.
We note that within the laned stripe phase, there is
an extended region where the system has stripe-like structure,
but the particles within each stripe are strongly fluctuating.
For values of $\rho$ higher than what we consider in this work,
it is likely that a uniformly jammed state would appear that would transition
into 
a uniformly laned state at high drives.
In Fig.~\ref{fig:18}(a), we illustrate the jammed labyrinth phase at
$\rho = 1.235$ and $F_D = 0.0$.
Figure~\ref{fig:18}(b) shows the disordered moving liquid phase
at $F_D = 0.2$ near the unjamming point,
while Fig.~\ref{fig:18}(c) shows the
alignment in the moving liquid phase for intermediate drives
of $F_D = 0.8$.
The high drive laned stripe state appears in
Fig.~\ref{fig:18}(d) at $F_D = 3.0$.

\begin{figure}
\includegraphics[width=\columnwidth]{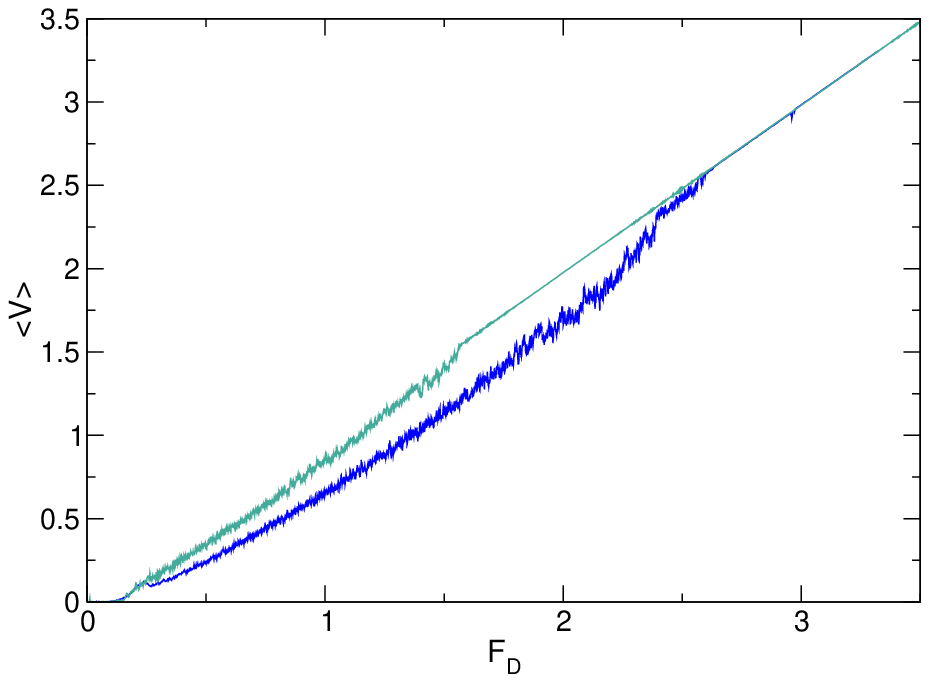} 
\caption{$\langle V \rangle$ vs $F_D$ for the system
from Fig.~\ref{fig:17} with $B=2.3$ at $\rho=0.47$ (green)
and $\rho=1.29$ (blue).
}
\label{fig:19M}
\end{figure}

In Fig.~\ref{fig:19M}, we plot
$\langle V \rangle$ versus $F_D$ for the $B=2.3$ system from Fig.~\ref{fig:17}
at $\rho=0.47$ and $\rho=1.29$.
For both densities,
the onset of the ordered laned stripe
phase is visible as a drop in the velocity fluctuations.
The disordered moving liquid regime is smaller for $\rho=0.47$ compared
to $\rho=1.29$,
and the ordering into the
laned stripe phase occurs at lower drives.

\begin{figure}
\includegraphics[width=\columnwidth]{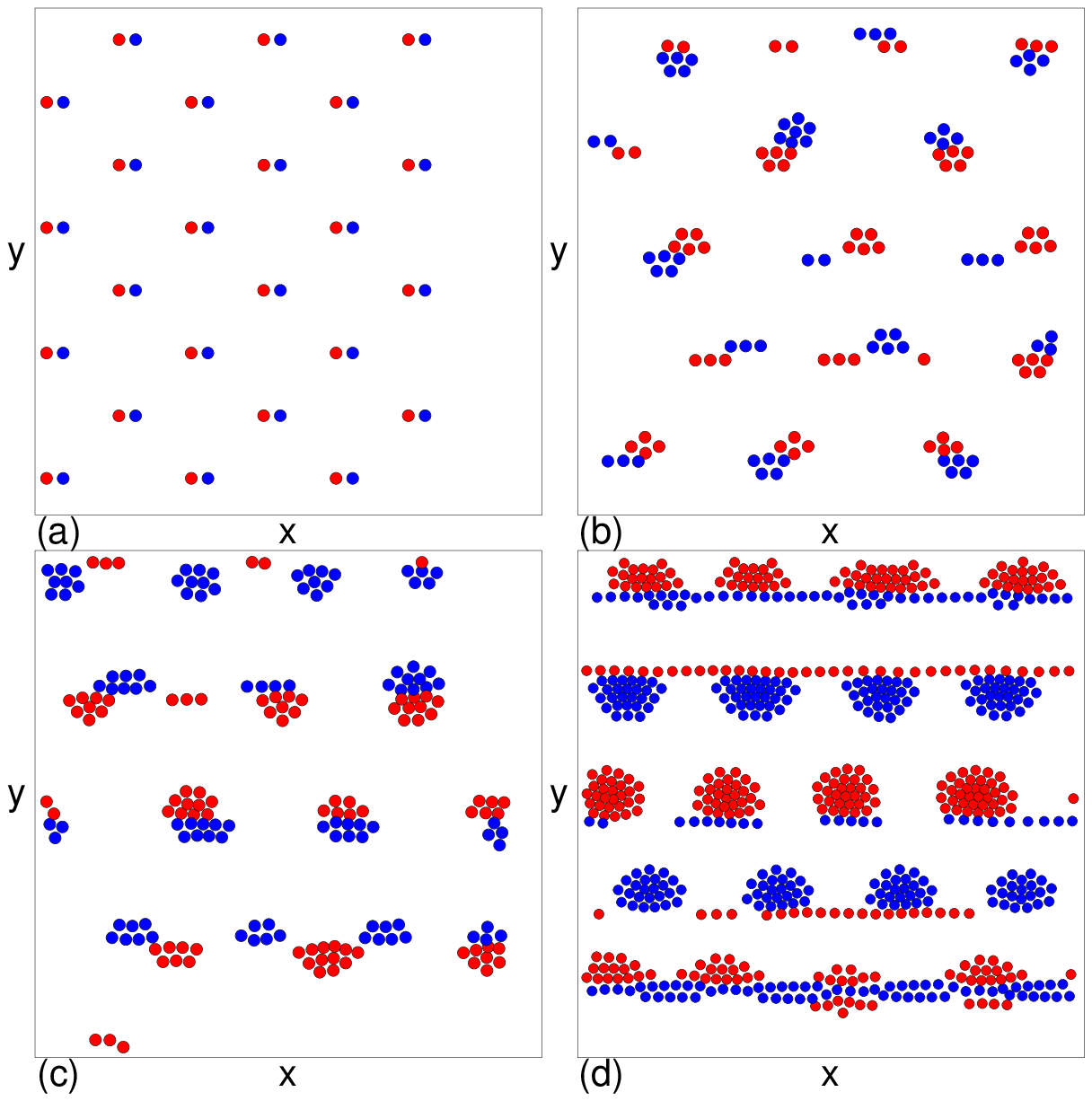}
\caption{Particle positions for the system from Fig.~\ref{fig:17}
with $B = 2.3$.
Red (blue) particles are driven in the negative (positive) $x$ direction.
(a) Jammed dimer phase at $\rho = 0.044$ and $F_D = 0.05$.
(b) Laned bubble phase at $\rho = 0.088$ and $F_D = 0.2$.
(c) Laned bubble phase at $\rho = 0.264$ and $F_D = 1.5$,
at the border with the laned stripe phase.
(d) Laned stripe state at $\rho = 0.529$ and $F_D = 2.0$.
}
\label{fig:19}
\end{figure}

We find some interesting ordered jammed and
laned states at low densities in the $B=2.3$ system from
Fig.~\ref{fig:17}.
In Fig.~\ref{fig:19}(a), we show that the jammed phase at $\rho = 0.044$
consists of polarized dimers.
An example of the moving laned bubble state at $F_D=1.5$ appears
in Fig.~\ref{fig:19}(b) for $\rho=0.88$, where the bubbles are quite
fragmentary.
For $\rho=0.264$ and $F_D=1.5$, at the border separating the laned
bubble and laned stripe states, Fig.~\ref{fig:19}(c) shows that the
bubbles have become more cohesive and are beginning to elongate.
Figure~\ref{fig:19}(d) shows the laned
stripe state at $\rho = 0.529$ and $F_D = 2.0$.

\begin{figure}
\includegraphics[width=\columnwidth]{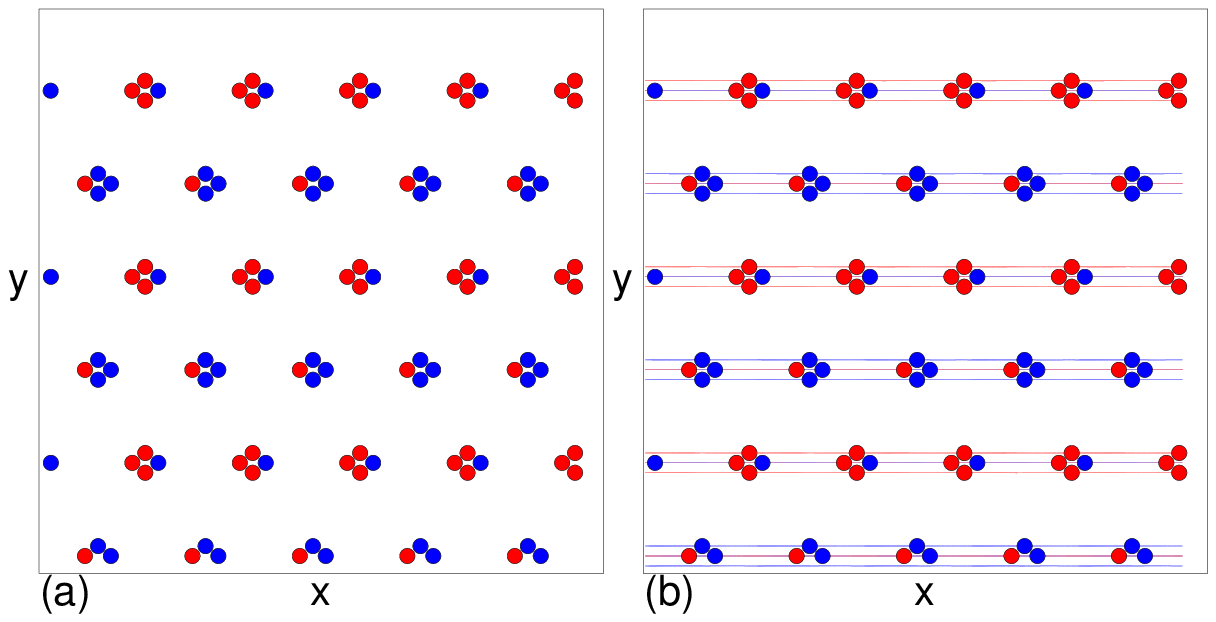}
\caption{Particle positions in the
laned plastic bubble state at $B = 2.4$, $\rho = 0.09$, and
$F_D = 0.15$.
Red (blue) particles are driven in the negative (positive) $x$ direction.
(b) The same system with lines indicating the particle trajectories,
where light red (light blue) lines illustrate the motion of particles driven
in the negative (positive) $x$ direction.
The bubbles each contain three particles of one species and
one particle of the other species, and the resulting force imbalance
leads to ordered flow.
}
\label{fig:20}
\end{figure}

Another unusual ordered bubble state we observed appears
in Fig.~\ref{fig:20}(a), where we show the particle configurations at
$\rho = 0.09$, $B = 2.4$, and $F_D = 0.15$.
Figure~\ref{fig:20}(b) shows the corresponding particle trajectories.
Each bubble captures four particles that are divided unevenly: there are three
particles of one species, and one particle of the other species. The
resulting force imbalance on each bubble produces motion, and
the
system has organized into alternating rows
containing bubbles of equal composition.
This is an example of a laned plastic bubble state.

\begin{figure}
\includegraphics[width=\columnwidth]{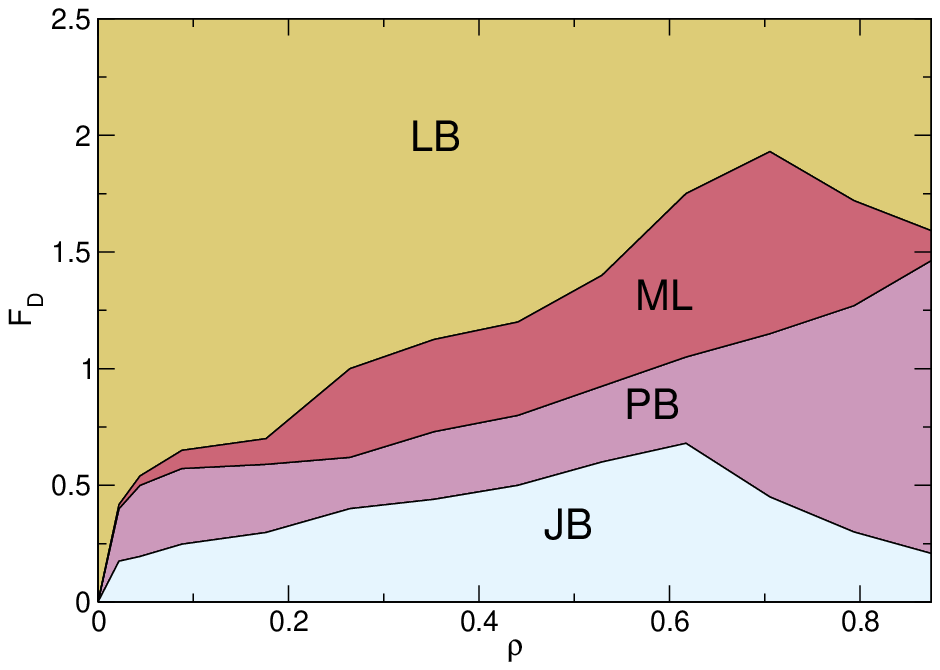}
\caption{Dynamic phase diagram as a function of $F_D$ vs
$\rho$ at $B = 2.8$, highlighting the extended plastic bubble phase.
JB: jammed bubble phase (pale blue),
PB: plastic bubble phase (purple),
ML: disordered moving liquid phase (red),
and LB: laned bubble phase (yellow).
}
\label{fig:21}
\end{figure}

In Fig.~\ref{fig:21}, we show a dynamic phase diagram
as a function of $F_D$ versus $\rho$ at $B = 2.8$.
In this case, $B$ is large enough to produce
an extended window of plastic bubble phase.
As $\rho$ increases, the size of the bubbles increases,
and the critical unjamming force $F_c$ also increases.
When $\rho>0.6$, however, $F_c$ begins to decrease with increasing
$\rho$, but since the onset of the moving liquid phase continues to
shift to higher $F_D$ with increasing $\rho$, the width of the
plastic bubble phase increases as the density becomes higher.

\begin{figure}
\includegraphics[width=\columnwidth]{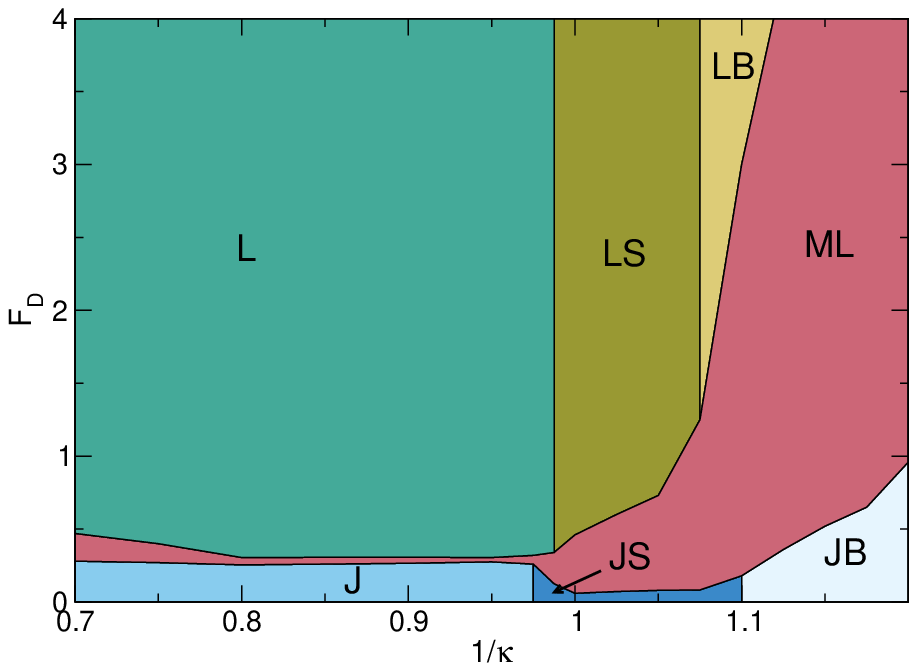}
\caption{Dynamic phase diagram as a function of
$F_D$ vs $1/\kappa$ for a system with $B = 2.0$ and $\rho = 0.441$.
JB: jammed bubble phase (pale blue),
JS: jammed stripe phase (dark blue),
J: jammed uniform phase (medium blue),
ML: disordered moving liquid phase (red),
LB: laned bubble phase (yellow),
LS: laned stripe phase (olive green),
and L: uniform laned state (green).
The critical unjamming force shows a dip in the stripe phase.
}
\label{fig:22}
\end{figure}

We have also considered a system in which we fix $B = 2.0$ and $\rho = 0.441$
while varying $1/\kappa$.
In Fig.~\ref{fig:22}, we plot the resulting
dynamic phase diagram as a function of $F_D$ versus $1/\kappa$.
The uniform jammed state that appears for $0.7 < 1/\kappa < 0.975$
unjams into a moving liquid state and then
transitions at higher drives to a uniform laned state.
For $0.975 \leq 1/\kappa < 1.1$, we find a jammed stripe state,
and for $1/\kappa > 1.1$, there is a jammed bubble state.
The critical unjamming threshold drops at
the transition from the uniform jammed state to the
jammed stripe state, and then begins to increase again
in the bubble phase.
The jammed stripe state unjams into a disordered moving liquid state and
then organizes into a laned stripe state for lower $1/\kappa$ and
into a laned bubble state for larger $1/\kappa$.

\begin{figure}
\includegraphics[width=\columnwidth]{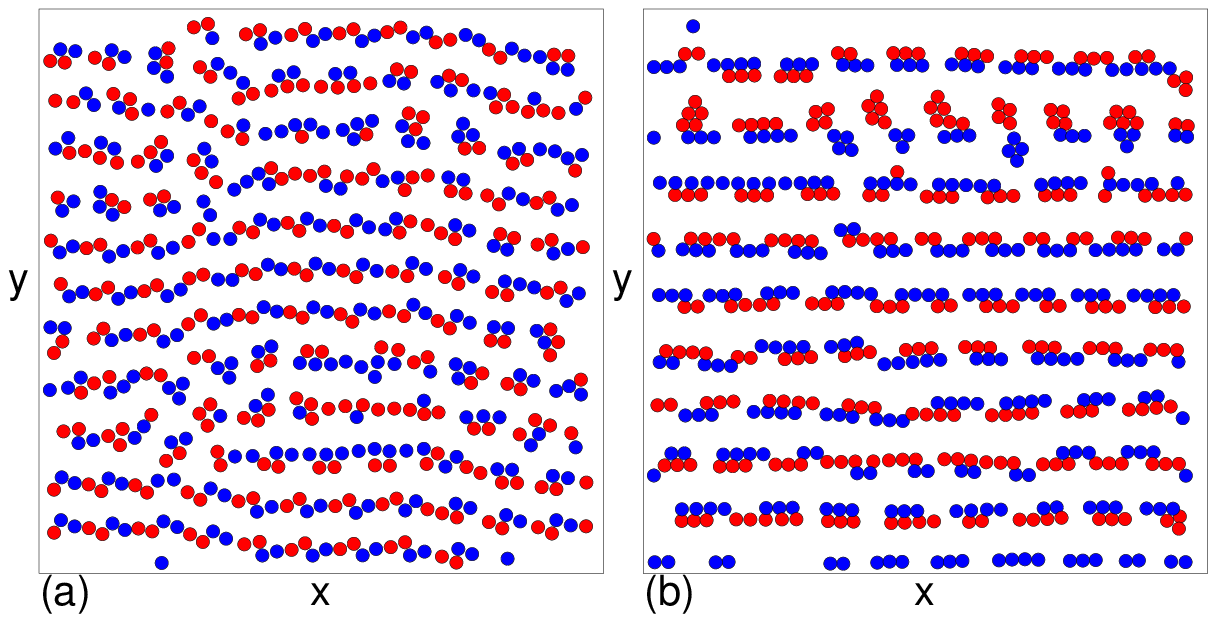}
\caption{Particle positions for a system with
$\rho = 0.441$, $1/\kappa = 0.52$, and $B = 6.0$.
Red (blue) particles are driven in the negative (positive) $x$ direction.  
(a) Jammed stripe state at $F_D = 0.05$.
(b) Moving laned stripe state at $F_D = 1.0$.
}
\label{fig:23}
\end{figure}

There are also a number of variations of the stripe and bubble
states that can arise in our system for varied $B$ and $1/\kappa$.
An example of this is shown in Fig.~\ref{fig:23}(a) for $B = 6.0$,
$\rho = 0.441$, and $1/\kappa = 0.52$ at $F_D=0.05$,
where a jammed stripe state appears
that has some local dimer structuring.
Figure~\ref{fig:23}(b) shows that at a higher drive of $F_D=1.0$, this
system forms a laned stripe state that resembles
the laned stripes found in the $\kappa=1.0$ system, but with
lanes that are greatly reduced in width.

\section{Discussion}

We considered a specific type of SALR interaction;
however, there are many other interactions that give similar patterns,
and we expect that such interactions would produce effects similar to
those described above.
A possible difference is that the jamming state may occur at much lower
density and could be much more strongly heterogeneous
for two-step repulsive potentials with finite range.
In certain cases, it may be easier in experiment to apply ac rather
than dc driving.
Since the patterns formed by our system have
multiple length scales,
ac driving could produce additional effects not found for single-scale strictly
repulsive interaction potentials. For example,
there could be one type of behavior at
ac amplitudes where the particle motion is on roughly the same scale as the
spacing of individual particles within a mesophase, but different behavior
could appear at higher ac amplitudes when the motion reaches the same scale
as the mesophase, such as the distance between stripes or bubbles.
We focused on the case of 50:50 mixtures of driven particles,
but if one species is more prevalent than the other, we expect that
some of the phases we observe could be modified.
We did not consider thermal fluctuations, but we expect that if the fluctuations
are weak, the same phases would arise.
At higher temperatures, it is likely that
the laned stripe state would be the most susceptible 
to transitioning into a disordered fluctuating quasi-one-dimensional state.
Beyond soft matter systems, our results could be relevant for certain
types of pedestrian flows, where instead of individual particles driven
in opposite directions, there are groups with some associations,
leading to an effective short range attraction between people.

\section{Summary}

We have examined the laning and dynamic phases for
an assembly of particles with competing short-range attraction
and long-range repulsion where half of the particles are driven in a direction
opposite to the other half.
In the absence of driving, this system
forms crystal, stripe, and bubble states depending on the density and
the ratio of attraction to repulsion.
In previous work for oppositely driven particles with single-scale
repulsive interactions,
it was shown that 
a laning state forms at high drives and a disordered state
appears at low drives.
For the system with competing interactions, we find
a much wider variety of states.
At high densities or when the repulsive term dominates,
as the drive increases the system passes through a uniform jammed state, a
disordered moving liquid state, and a high-drive laned state.
The transitions between these states produce signatures
in the velocity-force curve, differential mobility,
and density of topological defects.
For intermediate ratios of attraction to repulsion,
the system forms a jammed state that breaks up
near the unjamming transition.
At higher drives, there is a fluid-like stripe state
and then a laned stripe state where
each stripe contains well-defined lanes of oppositely moving particles.
For stronger attraction, the system forms a jammed bubble state.
As the drive increases, the particles remain jammed but rearrange
and segregate inside the bubbles to form
a strongly polarized elongated bubble state just before unjamming.
The system unjams into a fluctuating state or a plastic bubble state,
depending on the ability of the particles to jump
from bubble to bubble.
A variety of laned bubble states can appear,
including bubbles that move through each other
or a more fully segregated state where each lane contains
bubbles of a single particle species.
The unjamming threshold force
is strongly non-monotonic,
passing through a minimum in the stripe regime
and increasing in the bubble regime.
We map out the evolution of the dynamic
phases as a function of the ratio of attraction to repulsion,
particle density, screening length, and drive.

\begin{acknowledgements}
We gratefully acknowledge the support of the U.S. Department of
Energy through the LANL/LDRD program for this work.
This work was supported by the US Department of Energy through
the Los Alamos National Laboratory.  Los Alamos National Laboratory is
operated by Triad National Security, LLC, for the National Nuclear Security
Administration of the U. S. Department of Energy (Contract No. 892333218NCA000001).
\end{acknowledgements}

\bibliography{mybib}

\begin{thebibliography}{63}%
\makeatletter
\providecommand \@ifxundefined [1]{%
 \@ifx{#1\undefined}
}%
\providecommand \@ifnum [1]{%
 \ifnum #1\expandafter \@firstoftwo
 \else \expandafter \@secondoftwo
 \fi
}%
\providecommand \@ifx [1]{%
 \ifx #1\expandafter \@firstoftwo
 \else \expandafter \@secondoftwo
 \fi
}%
\providecommand \natexlab [1]{#1}%
\providecommand \enquote  [1]{``#1''}%
\providecommand \bibnamefont  [1]{#1}%
\providecommand \bibfnamefont [1]{#1}%
\providecommand \citenamefont [1]{#1}%
\providecommand \href@noop [0]{\@secondoftwo}%
\providecommand \href [0]{\begingroup \@sanitize@url \@href}%
\providecommand \@href[1]{\@@startlink{#1}\@@href}%
\providecommand \@@href[1]{\endgroup#1\@@endlink}%
\providecommand \@sanitize@url [0]{\catcode `\\12\catcode `\$12\catcode
  `\&12\catcode `\#12\catcode `\^12\catcode `\_12\catcode `\%12\relax}%
\providecommand \@@startlink[1]{}%
\providecommand \@@endlink[0]{}%
\providecommand \url  [0]{\begingroup\@sanitize@url \@url }%
\providecommand \@url [1]{\endgroup\@href {#1}{\urlprefix }}%
\providecommand \urlprefix  [0]{URL }%
\providecommand \Eprint [0]{\href }%
\providecommand \doibase [0]{https://doi.org/}%
\providecommand \selectlanguage [0]{\@gobble}%
\providecommand \bibinfo  [0]{\@secondoftwo}%
\providecommand \bibfield  [0]{\@secondoftwo}%
\providecommand \translation [1]{[#1]}%
\providecommand \BibitemOpen [0]{}%
\providecommand \bibitemStop [0]{}%
\providecommand \bibitemNoStop [0]{.\EOS\space}%
\providecommand \EOS [0]{\spacefactor3000\relax}%
\providecommand \BibitemShut  [1]{\csname bibitem#1\endcsname}%
\let\auto@bib@innerbib\@empty
\bibitem [{\citenamefont {Helbing}\ \emph {et~al.}(2000)\citenamefont
  {Helbing}, \citenamefont {Farkas},\ and\ \citenamefont {Vicsek}}]{Helbing00}%
  \BibitemOpen
  \bibfield  {author} {\bibinfo {author} {\bibfnamefont {D.}~\bibnamefont
  {Helbing}}, \bibinfo {author} {\bibfnamefont {I.~J.}\ \bibnamefont
  {Farkas}},\ and\ \bibinfo {author} {\bibfnamefont {T.}~\bibnamefont
  {Vicsek}},\ }\bibfield  {title} {\bibinfo {title} {Freezing by heating in a
  driven mesoscopic system},\ }\href
  {https://doi.org/10.1103/PhysRevLett.84.1240} {\bibfield  {journal} {\bibinfo
   {journal} {Phys. Rev. Lett.}\ }\textbf {\bibinfo {volume} {84}},\ \bibinfo
  {pages} {1240} (\bibinfo {year} {2000})}\BibitemShut {NoStop}%
\bibitem [{\citenamefont {Dzubiella}\ \emph {et~al.}(2002)\citenamefont
  {Dzubiella}, \citenamefont {Hoffmann},\ and\ \citenamefont
  {L\"owen}}]{Dzubiella02a}%
  \BibitemOpen
  \bibfield  {author} {\bibinfo {author} {\bibfnamefont {J.}~\bibnamefont
  {Dzubiella}}, \bibinfo {author} {\bibfnamefont {G.~P.}\ \bibnamefont
  {Hoffmann}},\ and\ \bibinfo {author} {\bibfnamefont {H.}~\bibnamefont
  {L\"owen}},\ }\bibfield  {title} {\bibinfo {title} {Lane formation in
  colloidal mixtures driven by an external field},\ }\href
  {https://doi.org/10.1103/PhysRevE.65.021402} {\bibfield  {journal} {\bibinfo
  {journal} {Phys. Rev. E}\ }\textbf {\bibinfo {volume} {65}},\ \bibinfo
  {pages} {021402} (\bibinfo {year} {2002})}\BibitemShut {NoStop}%
\bibitem [{\citenamefont {Chakrabarti}\ \emph {et~al.}(2004)\citenamefont
  {Chakrabarti}, \citenamefont {Dzubiella},\ and\ \citenamefont
  {L\"owen}}]{Chakrabarti04}%
  \BibitemOpen
  \bibfield  {author} {\bibinfo {author} {\bibfnamefont {J.}~\bibnamefont
  {Chakrabarti}}, \bibinfo {author} {\bibfnamefont {J.}~\bibnamefont
  {Dzubiella}},\ and\ \bibinfo {author} {\bibfnamefont {H.}~\bibnamefont
  {L\"owen}},\ }\bibfield  {title} {\bibinfo {title} {Reentrance effect in the
  lane formation of driven colloids},\ }\href
  {https://doi.org/10.1103/PhysRevE.70.012401} {\bibfield  {journal} {\bibinfo
  {journal} {Phys. Rev. E}\ }\textbf {\bibinfo {volume} {70}},\ \bibinfo
  {pages} {012401} (\bibinfo {year} {2004})}\BibitemShut {NoStop}%
\bibitem [{\citenamefont {Glanz}\ and\ \citenamefont {L{\"
  o}wen}(2012)}]{Glanz12}%
  \BibitemOpen
  \bibfield  {author} {\bibinfo {author} {\bibfnamefont {T.}~\bibnamefont
  {Glanz}}\ and\ \bibinfo {author} {\bibfnamefont {H.}~\bibnamefont {L{\"
  o}wen}},\ }\bibfield  {title} {\bibinfo {title} {The nature of the laning
  transition in two dimensions},\ }\href
  {https://doi.org/10.1088/0953-8984/24/46/464114} {\bibfield  {journal}
  {\bibinfo  {journal} {J. Phys: Condens. Matter}\ }\textbf {\bibinfo {volume}
  {24}},\ \bibinfo {pages} {464114} (\bibinfo {year} {2012})}\BibitemShut
  {NoStop}%
\bibitem [{\citenamefont {Ikeda}\ \emph {et~al.}(2012)\citenamefont {Ikeda},
  \citenamefont {Wada},\ and\ \citenamefont {Hayakawa}}]{Ikeda12}%
  \BibitemOpen
  \bibfield  {author} {\bibinfo {author} {\bibfnamefont {M.}~\bibnamefont
  {Ikeda}}, \bibinfo {author} {\bibfnamefont {H.}~\bibnamefont {Wada}},\ and\
  \bibinfo {author} {\bibfnamefont {H.}~\bibnamefont {Hayakawa}},\ }\bibfield
  {title} {\bibinfo {title} {Instabilities and turbulence-like dynamics in an
  oppositely driven binary particle mixture},\ }\href
  {https://doi.org/10.1209/0295-5075/99/68005} {\bibfield  {journal} {\bibinfo
  {journal} {EPL}\ }\textbf {\bibinfo {volume} {99}},\ \bibinfo {pages} {68005}
  (\bibinfo {year} {2012})}\BibitemShut {NoStop}%
\bibitem [{\citenamefont {Klymko}\ \emph {et~al.}(2016)\citenamefont {Klymko},
  \citenamefont {Geissler},\ and\ \citenamefont {Whitelam}}]{Klymko16}%
  \BibitemOpen
  \bibfield  {author} {\bibinfo {author} {\bibfnamefont {K.}~\bibnamefont
  {Klymko}}, \bibinfo {author} {\bibfnamefont {P.~L.}\ \bibnamefont
  {Geissler}},\ and\ \bibinfo {author} {\bibfnamefont {S.}~\bibnamefont
  {Whitelam}},\ }\bibfield  {title} {\bibinfo {title} {Microscopic origin and
  macroscopic implications of lane formation in mixtures of oppositely driven
  particles},\ }\href {https://doi.org/10.1103/PhysRevE.94.022608} {\bibfield
  {journal} {\bibinfo  {journal} {Phys. Rev. E}\ }\textbf {\bibinfo {volume}
  {94}},\ \bibinfo {pages} {022608} (\bibinfo {year} {2016})}\BibitemShut
  {NoStop}%
\bibitem [{\citenamefont {Poncet}\ \emph {et~al.}(2017)\citenamefont {Poncet},
  \citenamefont {B\'enichou}, \citenamefont {D\'emery},\ and\ \citenamefont
  {Oshanin}}]{Poncet17}%
  \BibitemOpen
  \bibfield  {author} {\bibinfo {author} {\bibfnamefont {A.}~\bibnamefont
  {Poncet}}, \bibinfo {author} {\bibfnamefont {O.}~\bibnamefont {B\'enichou}},
  \bibinfo {author} {\bibfnamefont {V.}~\bibnamefont {D\'emery}},\ and\
  \bibinfo {author} {\bibfnamefont {G.}~\bibnamefont {Oshanin}},\ }\bibfield
  {title} {\bibinfo {title} {Universal long ranged correlations in driven
  binary mixtures},\ }\href {https://doi.org/10.1103/PhysRevLett.118.118002}
  {\bibfield  {journal} {\bibinfo  {journal} {Phys. Rev. Lett.}\ }\textbf
  {\bibinfo {volume} {118}},\ \bibinfo {pages} {118002} (\bibinfo {year}
  {2017})}\BibitemShut {NoStop}%
\bibitem [{\citenamefont {Isele}\ \emph {et~al.}(2023)\citenamefont {Isele},
  \citenamefont {Hofmann}, \citenamefont {Erbe}, \citenamefont {Leiderer},\
  and\ \citenamefont {Nielaba}}]{Isele23}%
  \BibitemOpen
  \bibfield  {author} {\bibinfo {author} {\bibfnamefont {M.}~\bibnamefont
  {Isele}}, \bibinfo {author} {\bibfnamefont {K.}~\bibnamefont {Hofmann}},
  \bibinfo {author} {\bibfnamefont {A.}~\bibnamefont {Erbe}}, \bibinfo {author}
  {\bibfnamefont {P.}~\bibnamefont {Leiderer}},\ and\ \bibinfo {author}
  {\bibfnamefont {P.}~\bibnamefont {Nielaba}},\ }\bibfield  {title} {\bibinfo
  {title} {Lane formation of colloidal particles driven in parallel by
  gravity},\ }\href {https://doi.org/10.1103/PhysRevE.108.034607} {\bibfield
  {journal} {\bibinfo  {journal} {Phys. Rev. E}\ }\textbf {\bibinfo {volume}
  {108}},\ \bibinfo {pages} {034607} (\bibinfo {year} {2023})}\BibitemShut
  {NoStop}%
\bibitem [{\citenamefont {Yu}\ and\ \citenamefont {Jack}(2024)}]{Yu24}%
  \BibitemOpen
  \bibfield  {author} {\bibinfo {author} {\bibfnamefont {H.}~\bibnamefont
  {Yu}}\ and\ \bibinfo {author} {\bibfnamefont {R.~L.}\ \bibnamefont {Jack}},\
  }\bibfield  {title} {\bibinfo {title} {Competition between lanes and
  transient jammed clusters in driven binary mixtures},\ }\href
  {https://doi.org/10.1103/PhysRevE.109.024123} {\bibfield  {journal} {\bibinfo
   {journal} {Phys. Rev. E}\ }\textbf {\bibinfo {volume} {109}},\ \bibinfo
  {pages} {024123} (\bibinfo {year} {2024})}\BibitemShut {NoStop}%
\bibitem [{\citenamefont {Leunissen}\ \emph {et~al.}(2005)\citenamefont
  {Leunissen}, \citenamefont {Christova}, \citenamefont {Hynninen},
  \citenamefont {Royall}, \citenamefont {Campbell}, \citenamefont {Imhof},
  \citenamefont {Dijkstra}, \citenamefont {van Roij},\ and\ \citenamefont {van
  Blaaderen}}]{Leunissen05}%
  \BibitemOpen
  \bibfield  {author} {\bibinfo {author} {\bibfnamefont {M.~E.}\ \bibnamefont
  {Leunissen}}, \bibinfo {author} {\bibfnamefont {C.~G.}\ \bibnamefont
  {Christova}}, \bibinfo {author} {\bibfnamefont {A.~P.}\ \bibnamefont
  {Hynninen}}, \bibinfo {author} {\bibfnamefont {C.~P.}\ \bibnamefont
  {Royall}}, \bibinfo {author} {\bibfnamefont {A.~I.}\ \bibnamefont
  {Campbell}}, \bibinfo {author} {\bibfnamefont {A.}~\bibnamefont {Imhof}},
  \bibinfo {author} {\bibfnamefont {M.}~\bibnamefont {Dijkstra}}, \bibinfo
  {author} {\bibfnamefont {R.}~\bibnamefont {van Roij}},\ and\ \bibinfo
  {author} {\bibfnamefont {A.}~\bibnamefont {van Blaaderen}},\ }\bibfield
  {title} {\bibinfo {title} {Ionic colloidal crystals of oppositely charged
  particles},\ }\href {https://doi.org/10.1038/nature03946} {\bibfield
  {journal} {\bibinfo  {journal} {Nature (London)}\ }\textbf {\bibinfo {volume}
  {437}},\ \bibinfo {pages} {235} (\bibinfo {year} {2005})}\BibitemShut
  {NoStop}%
\bibitem [{\citenamefont {Rex}\ and\ \citenamefont {L\"owen}(2007)}]{Rex07}%
  \BibitemOpen
  \bibfield  {author} {\bibinfo {author} {\bibfnamefont {M.}~\bibnamefont
  {Rex}}\ and\ \bibinfo {author} {\bibfnamefont {H.}~\bibnamefont {L\"owen}},\
  }\bibfield  {title} {\bibinfo {title} {Lane formation in oppositely charged
  colloids driven by an electric field: Chaining and two-dimensional
  crystallization},\ }\href {https://doi.org/10.1103/PhysRevE.75.051402}
  {\bibfield  {journal} {\bibinfo  {journal} {Phys. Rev. E}\ }\textbf {\bibinfo
  {volume} {75}},\ \bibinfo {pages} {051402} (\bibinfo {year}
  {2007})}\BibitemShut {NoStop}%
\bibitem [{\citenamefont {L{\" o}wen}(2010)}]{Lowen10}%
  \BibitemOpen
  \bibfield  {author} {\bibinfo {author} {\bibfnamefont {H.}~\bibnamefont {L{\"
  o}wen}},\ }\bibfield  {title} {\bibinfo {title} {Particle-resolved
  instabilities in colloidal dispersions},\ }\href
  {https://doi.org/10.1039/B923685F} {\bibfield  {journal} {\bibinfo  {journal}
  {Soft Matter}\ }\textbf {\bibinfo {volume} {6}},\ \bibinfo {pages} {3133}
  (\bibinfo {year} {2010})}\BibitemShut {NoStop}%
\bibitem [{\citenamefont {Vissers}\ \emph {et~al.}(2011)\citenamefont
  {Vissers}, \citenamefont {van Blaaderen},\ and\ \citenamefont
  {Imhof}}]{Vissers11a}%
  \BibitemOpen
  \bibfield  {author} {\bibinfo {author} {\bibfnamefont {T.}~\bibnamefont
  {Vissers}}, \bibinfo {author} {\bibfnamefont {A.}~\bibnamefont {van
  Blaaderen}},\ and\ \bibinfo {author} {\bibfnamefont {A.}~\bibnamefont
  {Imhof}},\ }\bibfield  {title} {\bibinfo {title} {Band formation in mixtures
  of oppositely charged colloids driven by an ac electric field},\ }\href
  {https://doi.org/10.1103/PhysRevLett.106.228303} {\bibfield  {journal}
  {\bibinfo  {journal} {Phys. Rev. Lett.}\ }\textbf {\bibinfo {volume} {106}},\
  \bibinfo {pages} {228303} (\bibinfo {year} {2011})}\BibitemShut {NoStop}%
\bibitem [{\citenamefont {Geigenfeind}\ \emph {et~al.}(2020)\citenamefont
  {Geigenfeind}, \citenamefont {de~las Heras},\ and\ \citenamefont
  {Schmidt}}]{Geigeneind20}%
  \BibitemOpen
  \bibfield  {author} {\bibinfo {author} {\bibfnamefont {T.}~\bibnamefont
  {Geigenfeind}}, \bibinfo {author} {\bibfnamefont {D.}~\bibnamefont {de~las
  Heras}},\ and\ \bibinfo {author} {\bibfnamefont {M.}~\bibnamefont
  {Schmidt}},\ }\bibfield  {title} {\bibinfo {title} {Superadiabatic demixing
  in nonequilibrium colloids},\ }\href
  {https://doi.org/10.1038/s42005-020-0287-5} {\bibfield  {journal} {\bibinfo
  {journal} {Commun. Phys.}\ }\textbf {\bibinfo {volume} {3}},\ \bibinfo
  {pages} {23} (\bibinfo {year} {2020})}\BibitemShut {NoStop}%
\bibitem [{\citenamefont {S\"utterlin}\ \emph {et~al.}(2009)\citenamefont
  {S\"utterlin}, \citenamefont {Wysocki}, \citenamefont {Ivlev}, \citenamefont
  {R\"ath}, \citenamefont {Thomas}, \citenamefont {Rubin-Zuzic}, \citenamefont
  {Goedheer}, \citenamefont {Fortov}, \citenamefont {Lipaev}, \citenamefont
  {Molotkov}, \citenamefont {Petrov}, \citenamefont {Morfill},\ and\
  \citenamefont {L\"owen}}]{Sutterlin09}%
  \BibitemOpen
  \bibfield  {author} {\bibinfo {author} {\bibfnamefont {K.~R.}\ \bibnamefont
  {S\"utterlin}}, \bibinfo {author} {\bibfnamefont {A.}~\bibnamefont
  {Wysocki}}, \bibinfo {author} {\bibfnamefont {A.~V.}\ \bibnamefont {Ivlev}},
  \bibinfo {author} {\bibfnamefont {C.}~\bibnamefont {R\"ath}}, \bibinfo
  {author} {\bibfnamefont {H.~M.}\ \bibnamefont {Thomas}}, \bibinfo {author}
  {\bibfnamefont {M.}~\bibnamefont {Rubin-Zuzic}}, \bibinfo {author}
  {\bibfnamefont {W.~J.}\ \bibnamefont {Goedheer}}, \bibinfo {author}
  {\bibfnamefont {V.~E.}\ \bibnamefont {Fortov}}, \bibinfo {author}
  {\bibfnamefont {A.~M.}\ \bibnamefont {Lipaev}}, \bibinfo {author}
  {\bibfnamefont {V.~I.}\ \bibnamefont {Molotkov}}, \bibinfo {author}
  {\bibfnamefont {O.~F.}\ \bibnamefont {Petrov}}, \bibinfo {author}
  {\bibfnamefont {G.~E.}\ \bibnamefont {Morfill}},\ and\ \bibinfo {author}
  {\bibfnamefont {H.}~\bibnamefont {L\"owen}},\ }\bibfield  {title} {\bibinfo
  {title} {Dynamics of lane formation in driven binary complex plasmas},\
  }\href {https://doi.org/10.1103/PhysRevLett.102.085003} {\bibfield  {journal}
  {\bibinfo  {journal} {Phys. Rev. Lett.}\ }\textbf {\bibinfo {volume} {102}},\
  \bibinfo {pages} {085003} (\bibinfo {year} {2009})}\BibitemShut {NoStop}%
\bibitem [{\citenamefont {Vizarim}\ \emph {et~al.}(2025)\citenamefont
  {Vizarim}, \citenamefont {Souza}, \citenamefont {Reichhardt}, \citenamefont
  {Reichhardt}, \citenamefont {Venegas},\ and\ \citenamefont
  {B\'eron}}]{Vizarim25}%
  \BibitemOpen
  \bibfield  {author} {\bibinfo {author} {\bibfnamefont {N.~P.}\ \bibnamefont
  {Vizarim}}, \bibinfo {author} {\bibfnamefont {J.~C.~B.}\ \bibnamefont
  {Souza}}, \bibinfo {author} {\bibfnamefont {C.~J.~O.}\ \bibnamefont
  {Reichhardt}}, \bibinfo {author} {\bibfnamefont {C.}~\bibnamefont
  {Reichhardt}}, \bibinfo {author} {\bibfnamefont {P.~A.}\ \bibnamefont
  {Venegas}},\ and\ \bibinfo {author} {\bibfnamefont {F.}~\bibnamefont
  {B\'eron}},\ }\bibfield  {title} {\bibinfo {title} {Skyrmion-skyrmionium
  phase separation and laning transitions via spin-orbit torque currents},\
  }\href {https://doi.org/10.1103/zq91-42cc} {\bibfield  {journal} {\bibinfo
  {journal} {Phys. Rev. B}\ }\textbf {\bibinfo {volume} {111}},\ \bibinfo
  {pages} {214438} (\bibinfo {year} {2025})}\BibitemShut {NoStop}%
\bibitem [{\citenamefont {Reichhardt}\ and\ \citenamefont
  {Reichhardt}(2018)}]{Reichhardt18}%
  \BibitemOpen
  \bibfield  {author} {\bibinfo {author} {\bibfnamefont {C.}~\bibnamefont
  {Reichhardt}}\ and\ \bibinfo {author} {\bibfnamefont {C.~J.~O.}\ \bibnamefont
  {Reichhardt}},\ }\bibfield  {title} {\bibinfo {title} {Velocity force curves,
  laning, and jamming for oppositely driven disk systems},\ }\href
  {https://doi.org/10.1039/c7sm02162c} {\bibfield  {journal} {\bibinfo
  {journal} {Soft Matter}\ }\textbf {\bibinfo {volume} {14}},\ \bibinfo {pages}
  {490} (\bibinfo {year} {2018})}\BibitemShut {NoStop}%
\bibitem [{\citenamefont {Feliciani}\ and\ \citenamefont
  {Nishinari}(2016)}]{Feliciani16}%
  \BibitemOpen
  \bibfield  {author} {\bibinfo {author} {\bibfnamefont {C.}~\bibnamefont
  {Feliciani}}\ and\ \bibinfo {author} {\bibfnamefont {K.}~\bibnamefont
  {Nishinari}},\ }\bibfield  {title} {\bibinfo {title} {Empirical analysis of
  the lane formation process in bidirectional pedestrian flow},\ }\href
  {https://doi.org/10.1103/PhysRevE.94.032304} {\bibfield  {journal} {\bibinfo
  {journal} {Phys. Rev. E}\ }\textbf {\bibinfo {volume} {94}},\ \bibinfo
  {pages} {032304} (\bibinfo {year} {2016})}\BibitemShut {NoStop}%
\bibitem [{\citenamefont {Bacik}\ \emph {et~al.}(2023)\citenamefont {Bacik},
  \citenamefont {Bacik},\ and\ \citenamefont {Rogers}}]{Bacik23}%
  \BibitemOpen
  \bibfield  {author} {\bibinfo {author} {\bibfnamefont {K.~A.}\ \bibnamefont
  {Bacik}}, \bibinfo {author} {\bibfnamefont {B.~S.}\ \bibnamefont {Bacik}},\
  and\ \bibinfo {author} {\bibfnamefont {T.}~\bibnamefont {Rogers}},\
  }\bibfield  {title} {\bibinfo {title} {Lane nucleation in complex active
  flows},\ }\href {https://doi.org/10.1126/science.add8091} {\bibfield
  {journal} {\bibinfo  {journal} {Science}\ }\textbf {\bibinfo {volume}
  {379}},\ \bibinfo {pages} {923} (\bibinfo {year} {2023})}\BibitemShut
  {NoStop}%
\bibitem [{\citenamefont {Bain}\ and\ \citenamefont {Bartolo}(2017)}]{Bain17}%
  \BibitemOpen
  \bibfield  {author} {\bibinfo {author} {\bibfnamefont {N.}~\bibnamefont
  {Bain}}\ and\ \bibinfo {author} {\bibfnamefont {D.}~\bibnamefont {Bartolo}},\
  }\bibfield  {title} {\bibinfo {title} {Critical mingling and universal
  correlations in model binary active liquids},\ }\href
  {https://doi.org/10.1038/ncomms15969} {\bibfield  {journal} {\bibinfo
  {journal} {Nature Commun.}\ }\textbf {\bibinfo {volume} {8}},\ \bibinfo
  {pages} {15969} (\bibinfo {year} {2017})}\BibitemShut {NoStop}%
\bibitem [{\citenamefont {Reichhardt}\ \emph {et~al.}(2018)\citenamefont
  {Reichhardt}, \citenamefont {Thibault}, \citenamefont {Papanikolaou},\ and\
  \citenamefont {Reichhardt}}]{Reichhardt18b}%
  \BibitemOpen
  \bibfield  {author} {\bibinfo {author} {\bibfnamefont {C.}~\bibnamefont
  {Reichhardt}}, \bibinfo {author} {\bibfnamefont {J.}~\bibnamefont
  {Thibault}}, \bibinfo {author} {\bibfnamefont {S.}~\bibnamefont
  {Papanikolaou}},\ and\ \bibinfo {author} {\bibfnamefont {C.~J.~O.}\
  \bibnamefont {Reichhardt}},\ }\bibfield  {title} {\bibinfo {title} {Laning
  and clustering transitions in driven binary active matter systems},\ }\href
  {https://doi.org/10.1103/PhysRevE.98.022603} {\bibfield  {journal} {\bibinfo
  {journal} {Phys. Rev. E}\ }\textbf {\bibinfo {volume} {98}},\ \bibinfo
  {pages} {022603} (\bibinfo {year} {2018})}\BibitemShut {NoStop}%
\bibitem [{\citenamefont {Khelfa}\ \emph {et~al.}(2022)\citenamefont {Khelfa},
  \citenamefont {Korbmacher}, \citenamefont {Schadschneider},\ and\
  \citenamefont {Tordeux}}]{Khelfa22}%
  \BibitemOpen
  \bibfield  {author} {\bibinfo {author} {\bibfnamefont {B.}~\bibnamefont
  {Khelfa}}, \bibinfo {author} {\bibfnamefont {R.}~\bibnamefont {Korbmacher}},
  \bibinfo {author} {\bibfnamefont {A.}~\bibnamefont {Schadschneider}},\ and\
  \bibinfo {author} {\bibfnamefont {A.}~\bibnamefont {Tordeux}},\ }\bibfield
  {title} {\bibinfo {title} {Heterogeneity-induced lane and band formation in
  self-driven particle systems},\ }\href
  {https://doi.org/10.1038/s41598-022-08649-4} {\bibfield  {journal} {\bibinfo
  {journal} {Sci. Rep.}\ }\textbf {\bibinfo {volume} {12}},\ \bibinfo {pages}
  {4768} (\bibinfo {year} {2022})}\BibitemShut {NoStop}%
\bibitem [{\citenamefont {Glanz}\ \emph {et~al.}(2016)\citenamefont {Glanz},
  \citenamefont {Wittkowski},\ and\ \citenamefont {L\"owen}}]{Glanz16}%
  \BibitemOpen
  \bibfield  {author} {\bibinfo {author} {\bibfnamefont {T.}~\bibnamefont
  {Glanz}}, \bibinfo {author} {\bibfnamefont {R.}~\bibnamefont {Wittkowski}},\
  and\ \bibinfo {author} {\bibfnamefont {H.}~\bibnamefont {L\"owen}},\
  }\bibfield  {title} {\bibinfo {title} {Symmetry breaking in clogging for
  oppositely driven particles},\ }\href
  {https://doi.org/10.1103/PhysRevE.94.052606} {\bibfield  {journal} {\bibinfo
  {journal} {Phys. Rev. E}\ }\textbf {\bibinfo {volume} {94}},\ \bibinfo
  {pages} {052606} (\bibinfo {year} {2016})}\BibitemShut {NoStop}%
\bibitem [{\citenamefont {Wysocki}\ and\ \citenamefont
  {L\"owen}(2009)}]{Wysocki09}%
  \BibitemOpen
  \bibfield  {author} {\bibinfo {author} {\bibfnamefont {A.}~\bibnamefont
  {Wysocki}}\ and\ \bibinfo {author} {\bibfnamefont {H.}~\bibnamefont
  {L\"owen}},\ }\bibfield  {title} {\bibinfo {title} {Oscillatory driven
  colloidal binary mixtures: Axial segregation versus laning},\ }\href
  {https://doi.org/10.1103/PhysRevE.79.041408} {\bibfield  {journal} {\bibinfo
  {journal} {Phys. Rev. E}\ }\textbf {\bibinfo {volume} {79}},\ \bibinfo
  {pages} {041408} (\bibinfo {year} {2009})}\BibitemShut {NoStop}%
\bibitem [{\citenamefont {Li}\ \emph {et~al.}(2021)\citenamefont {Li},
  \citenamefont {Wang}, \citenamefont {Shi}, \citenamefont {Gao}, \citenamefont
  {Shi}, \citenamefont {Woodward},\ and\ \citenamefont {Forsman}}]{Li21}%
  \BibitemOpen
  \bibfield  {author} {\bibinfo {author} {\bibfnamefont {B.}~\bibnamefont
  {Li}}, \bibinfo {author} {\bibfnamefont {Y.-L.}\ \bibnamefont {Wang}},
  \bibinfo {author} {\bibfnamefont {G.}~\bibnamefont {Shi}}, \bibinfo {author}
  {\bibfnamefont {Y.}~\bibnamefont {Gao}}, \bibinfo {author} {\bibfnamefont
  {X.}~\bibnamefont {Shi}}, \bibinfo {author} {\bibfnamefont {C.~E.}\
  \bibnamefont {Woodward}},\ and\ \bibinfo {author} {\bibfnamefont
  {J.}~\bibnamefont {Forsman}},\ }\bibfield  {title} {\bibinfo {title} {Phase
  transitions of oppositely charged colloidal particles driven by alternating
  current electric field},\ }\href {https://doi.org/10.1021/acsnano.0c04095}
  {\bibfield  {journal} {\bibinfo  {journal} {ACS Nano}\ }\textbf {\bibinfo
  {volume} {15}},\ \bibinfo {pages} {2363} (\bibinfo {year}
  {2021})}\BibitemShut {NoStop}%
\bibitem [{\citenamefont {Reichhardt}\ and\ \citenamefont
  {Reichhardt}(2019)}]{Reichhardt19a}%
  \BibitemOpen
  \bibfield  {author} {\bibinfo {author} {\bibfnamefont {C.~J.~O.}\
  \bibnamefont {Reichhardt}}\ and\ \bibinfo {author} {\bibfnamefont
  {C.}~\bibnamefont {Reichhardt}},\ }\bibfield  {title} {\bibinfo {title}
  {Disordering, clustering, and laning transitions in particle systems with
  dispersion in the {M}agnus term},\ }\href
  {https://doi.org/10.1103/PhysRevE.99.012606} {\bibfield  {journal} {\bibinfo
  {journal} {Phys. Rev. E}\ }\textbf {\bibinfo {volume} {99}},\ \bibinfo
  {pages} {012606} (\bibinfo {year} {2019})}\BibitemShut {NoStop}%
\bibitem [{\citenamefont {W\"achtler}\ \emph {et~al.}(2016)\citenamefont
  {W\"achtler}, \citenamefont {Kogler},\ and\ \citenamefont
  {Klapp}}]{Wachtler16}%
  \BibitemOpen
  \bibfield  {author} {\bibinfo {author} {\bibfnamefont {C.~W.}\ \bibnamefont
  {W\"achtler}}, \bibinfo {author} {\bibfnamefont {F.}~\bibnamefont {Kogler}},\
  and\ \bibinfo {author} {\bibfnamefont {S.~H.~L.}\ \bibnamefont {Klapp}},\
  }\bibfield  {title} {\bibinfo {title} {Lane formation in a driven attractive
  fluid},\ }\href {https://doi.org/10.1103/PhysRevE.94.052603} {\bibfield
  {journal} {\bibinfo  {journal} {Phys. Rev. E}\ }\textbf {\bibinfo {volume}
  {94}},\ \bibinfo {pages} {052603} (\bibinfo {year} {2016})}\BibitemShut
  {NoStop}%
\bibitem [{\citenamefont {Kogler}\ and\ \citenamefont
  {Klapp}(2015)}]{Kogler15}%
  \BibitemOpen
  \bibfield  {author} {\bibinfo {author} {\bibfnamefont {F.}~\bibnamefont
  {Kogler}}\ and\ \bibinfo {author} {\bibfnamefont {S.~H.~L.}\ \bibnamefont
  {Klapp}},\ }\bibfield  {title} {\bibinfo {title} {Lane formation in a system
  of dipolar microswimmers},\ }\href
  {https://doi.org/10.1209/0295-5075/110/10004} {\bibfield  {journal} {\bibinfo
   {journal} {EPL}\ }\textbf {\bibinfo {volume} {110}},\ \bibinfo {pages}
  {10004} (\bibinfo {year} {2015})}\BibitemShut {NoStop}%
\bibitem [{\citenamefont {Seul}\ and\ \citenamefont {Andelman}(1995)}]{Seul95}%
  \BibitemOpen
  \bibfield  {author} {\bibinfo {author} {\bibfnamefont {M.}~\bibnamefont
  {Seul}}\ and\ \bibinfo {author} {\bibfnamefont {D.}~\bibnamefont
  {Andelman}},\ }\bibfield  {title} {\bibinfo {title} {Domain shapes and
  patterns - the phenomenology of modulated phases},\ }\href
  {https://doi.org/10.1126/science.267.5197.476} {\bibfield  {journal}
  {\bibinfo  {journal} {Science}\ }\textbf {\bibinfo {volume} {267}},\ \bibinfo
  {pages} {476} (\bibinfo {year} {1995})}\BibitemShut {NoStop}%
\bibitem [{\citenamefont {Stoycheva}\ and\ \citenamefont
  {Singer}(2000)}]{Stoycheva00}%
  \BibitemOpen
  \bibfield  {author} {\bibinfo {author} {\bibfnamefont {A.~D.}\ \bibnamefont
  {Stoycheva}}\ and\ \bibinfo {author} {\bibfnamefont {S.~J.}\ \bibnamefont
  {Singer}},\ }\bibfield  {title} {\bibinfo {title} {Stripe melting in a
  two-dimensional system with competing interactions},\ }\href
  {https://doi.org/10.1103/PhysRevLett.84.4657} {\bibfield  {journal} {\bibinfo
   {journal} {Phys. Rev. Lett.}\ }\textbf {\bibinfo {volume} {84}},\ \bibinfo
  {pages} {4657} (\bibinfo {year} {2000})}\BibitemShut {NoStop}%
\bibitem [{\citenamefont {Reichhardt}\ \emph {et~al.}(2003)\citenamefont
  {Reichhardt}, \citenamefont {Olson}, \citenamefont {Martin},\ and\
  \citenamefont {Bishop}}]{Reichhardt03}%
  \BibitemOpen
  \bibfield  {author} {\bibinfo {author} {\bibfnamefont {C.}~\bibnamefont
  {Reichhardt}}, \bibinfo {author} {\bibfnamefont {C.~J.}\ \bibnamefont
  {Olson}}, \bibinfo {author} {\bibfnamefont {I.}~\bibnamefont {Martin}},\ and\
  \bibinfo {author} {\bibfnamefont {A.~R.}\ \bibnamefont {Bishop}},\ }\bibfield
   {title} {\bibinfo {title} {Depinning and dynamics of systems with competing
  interactions in quenched disorder},\ }\href
  {https://doi.org/10.1209/epl/i2003-00222-0} {\bibfield  {journal} {\bibinfo
  {journal} {Europhys. Lett.}\ }\textbf {\bibinfo {volume} {61}},\ \bibinfo
  {pages} {221} (\bibinfo {year} {2003})}\BibitemShut {NoStop}%
\bibitem [{\citenamefont {Reichhardt}\ \emph {et~al.}(2004)\citenamefont
  {Reichhardt}, \citenamefont {Reichhardt}, \citenamefont {Martin},\ and\
  \citenamefont {Bishop}}]{Reichhardt04}%
  \BibitemOpen
  \bibfield  {author} {\bibinfo {author} {\bibfnamefont {C.~J.~O.}\
  \bibnamefont {Reichhardt}}, \bibinfo {author} {\bibfnamefont
  {C.}~\bibnamefont {Reichhardt}}, \bibinfo {author} {\bibfnamefont
  {I.}~\bibnamefont {Martin}},\ and\ \bibinfo {author} {\bibfnamefont {A.~R.}\
  \bibnamefont {Bishop}},\ }\bibfield  {title} {\bibinfo {title} {Dynamics and
  melting of stripes, crystals, and bubbles with quenched disorder},\ }\href
  {https://doi.org/10.1016/j.physd.2004.01.027} {\bibfield  {journal} {\bibinfo
   {journal} {Physica D}\ }\textbf {\bibinfo {volume} {193}},\ \bibinfo {pages}
  {303} (\bibinfo {year} {2004})}\BibitemShut {NoStop}%
\bibitem [{\citenamefont {Mossa}\ \emph {et~al.}(2004)\citenamefont {Mossa},
  \citenamefont {Sciortino}, \citenamefont {Tartaglia},\ and\ \citenamefont
  {Zaccarelli}}]{Mossa04}%
  \BibitemOpen
  \bibfield  {author} {\bibinfo {author} {\bibfnamefont {S.}~\bibnamefont
  {Mossa}}, \bibinfo {author} {\bibfnamefont {F.}~\bibnamefont {Sciortino}},
  \bibinfo {author} {\bibfnamefont {P.}~\bibnamefont {Tartaglia}},\ and\
  \bibinfo {author} {\bibfnamefont {E.}~\bibnamefont {Zaccarelli}},\ }\bibfield
   {title} {\bibinfo {title} {Ground-state clusters for short-range attractive
  and long-range repulsive potentials},\ }\href
  {https://doi.org/10.1021/la048554t} {\bibfield  {journal} {\bibinfo
  {journal} {Langmuir}\ }\textbf {\bibinfo {volume} {20}},\ \bibinfo {pages}
  {10756} (\bibinfo {year} {2004})}\BibitemShut {NoStop}%
\bibitem [{\citenamefont {Sciortino}\ \emph {et~al.}(2004)\citenamefont
  {Sciortino}, \citenamefont {Mossa}, \citenamefont {Zaccarelli},\ and\
  \citenamefont {Tartaglia}}]{Sciortino04}%
  \BibitemOpen
  \bibfield  {author} {\bibinfo {author} {\bibfnamefont {F.}~\bibnamefont
  {Sciortino}}, \bibinfo {author} {\bibfnamefont {S.}~\bibnamefont {Mossa}},
  \bibinfo {author} {\bibfnamefont {E.}~\bibnamefont {Zaccarelli}},\ and\
  \bibinfo {author} {\bibfnamefont {P.}~\bibnamefont {Tartaglia}},\ }\bibfield
  {title} {\bibinfo {title} {Equilibrium cluster phases and low-density
  arrested disordered states: The role of short-range attraction and long-range
  repulsion},\ }\href {https://doi.org/10.1103/PhysRevLett.93.055701}
  {\bibfield  {journal} {\bibinfo  {journal} {Phys. Rev. Lett.}\ }\textbf
  {\bibinfo {volume} {93}},\ \bibinfo {pages} {055701} (\bibinfo {year}
  {2004})}\BibitemShut {NoStop}%
\bibitem [{\citenamefont {Nelissen}\ \emph {et~al.}(2005)\citenamefont
  {Nelissen}, \citenamefont {Partoens},\ and\ \citenamefont
  {Peeters}}]{Nelissen05}%
  \BibitemOpen
  \bibfield  {author} {\bibinfo {author} {\bibfnamefont {K.}~\bibnamefont
  {Nelissen}}, \bibinfo {author} {\bibfnamefont {B.}~\bibnamefont {Partoens}},\
  and\ \bibinfo {author} {\bibfnamefont {F.~M.}\ \bibnamefont {Peeters}},\
  }\bibfield  {title} {\bibinfo {title} {Bubble, stripe, and ring phases in a
  two-dimensional cluster with competing interactions},\ }\href
  {https://doi.org/10.1103/PhysRevE.71.066204} {\bibfield  {journal} {\bibinfo
  {journal} {Phys. Rev. E}\ }\textbf {\bibinfo {volume} {71}},\ \bibinfo
  {pages} {066204} (\bibinfo {year} {2005})}\BibitemShut {NoStop}%
\bibitem [{\citenamefont {Liu}\ \emph {et~al.}(2008)\citenamefont {Liu},
  \citenamefont {Chew},\ and\ \citenamefont {Yu}}]{Liu08}%
  \BibitemOpen
  \bibfield  {author} {\bibinfo {author} {\bibfnamefont {Y.~H.}\ \bibnamefont
  {Liu}}, \bibinfo {author} {\bibfnamefont {L.~Y.}\ \bibnamefont {Chew}},\ and\
  \bibinfo {author} {\bibfnamefont {M.~Y.}\ \bibnamefont {Yu}},\ }\bibfield
  {title} {\bibinfo {title} {Self-assembly of complex structures in a
  two-dimensional system with competing interaction forces},\ }\href
  {https://doi.org/10.1103/PhysRevE.78.066405} {\bibfield  {journal} {\bibinfo
  {journal} {Phys. Rev. E}\ }\textbf {\bibinfo {volume} {78}},\ \bibinfo
  {pages} {066405} (\bibinfo {year} {2008})}\BibitemShut {NoStop}%
\bibitem [{\citenamefont {Olson~Reichhardt}\ \emph {et~al.}(2010)\citenamefont
  {Olson~Reichhardt}, \citenamefont {Reichhardt},\ and\ \citenamefont
  {Bishop}}]{Reichhardt10}%
  \BibitemOpen
  \bibfield  {author} {\bibinfo {author} {\bibfnamefont {C.~J.}\ \bibnamefont
  {Olson~Reichhardt}}, \bibinfo {author} {\bibfnamefont {C.}~\bibnamefont
  {Reichhardt}},\ and\ \bibinfo {author} {\bibfnamefont {A.~R.}\ \bibnamefont
  {Bishop}},\ }\bibfield  {title} {\bibinfo {title} {Structural transitions,
  melting, and intermediate phases for stripe- and clump-forming systems},\
  }\href {https://doi.org/10.1103/PhysRevE.82.041502} {\bibfield  {journal}
  {\bibinfo  {journal} {Phys. Rev. E}\ }\textbf {\bibinfo {volume} {82}},\
  \bibinfo {pages} {041502} (\bibinfo {year} {2010})}\BibitemShut {NoStop}%
\bibitem [{\citenamefont {McDermott}\ \emph {et~al.}(2014)\citenamefont
  {McDermott}, \citenamefont {Reichhardt},\ and\ \citenamefont
  {Reichhardt}}]{McDermott14}%
  \BibitemOpen
  \bibfield  {author} {\bibinfo {author} {\bibfnamefont {D.}~\bibnamefont
  {McDermott}}, \bibinfo {author} {\bibfnamefont {C.~J.~O.}\ \bibnamefont
  {Reichhardt}},\ and\ \bibinfo {author} {\bibfnamefont {C.}~\bibnamefont
  {Reichhardt}},\ }\bibfield  {title} {\bibinfo {title} {Stripe systems with
  competing interactions on quasi-one dimensional periodic substrates},\ }\href
  {https://doi.org/10.1039/c4sm01341g} {\bibfield  {journal} {\bibinfo
  {journal} {Soft Matter}\ }\textbf {\bibinfo {volume} {10}},\ \bibinfo {pages}
  {6332} (\bibinfo {year} {2014})}\BibitemShut {NoStop}%
\bibitem [{\citenamefont {Liu}\ and\ \citenamefont {Xi}(2019)}]{Liu19}%
  \BibitemOpen
  \bibfield  {author} {\bibinfo {author} {\bibfnamefont {Y.}~\bibnamefont
  {Liu}}\ and\ \bibinfo {author} {\bibfnamefont {Y.}~\bibnamefont {Xi}},\
  }\bibfield  {title} {\bibinfo {title} {Colloidal systems with a short-range
  attraction and long-range repulsion: phase diagrams, structures, and
  dynamics},\ }\href {https://doi.org/10.1016/j.cocis.2019.01.016} {\bibfield
  {journal} {\bibinfo  {journal} {Curr. Opin. Colloid Interf. Sci.}\ }\textbf
  {\bibinfo {volume} {19}},\ \bibinfo {pages} {123} (\bibinfo {year}
  {2019})}\BibitemShut {NoStop}%
\bibitem [{\citenamefont {Hooshanginejad}\ \emph {et~al.}(2024)\citenamefont
  {Hooshanginejad}, \citenamefont {Barotta}, \citenamefont {Spradlin},
  \citenamefont {Pucci}, \citenamefont {Hunt},\ and\ \citenamefont
  {Harris}}]{Hooshanginejad24}%
  \BibitemOpen
  \bibfield  {author} {\bibinfo {author} {\bibfnamefont {A.}~\bibnamefont
  {Hooshanginejad}}, \bibinfo {author} {\bibfnamefont {J.-W.}\ \bibnamefont
  {Barotta}}, \bibinfo {author} {\bibfnamefont {V.}~\bibnamefont {Spradlin}},
  \bibinfo {author} {\bibfnamefont {G.}~\bibnamefont {Pucci}}, \bibinfo
  {author} {\bibfnamefont {R.}~\bibnamefont {Hunt}},\ and\ \bibinfo {author}
  {\bibfnamefont {D.~M.}\ \bibnamefont {Harris}},\ }\bibfield  {title}
  {\bibinfo {title} {Interactions and pattern formation in a macroscopic
  magnetocapillary salr system of mermaid cereal},\ }\href
  {https://doi.org/10.1038/s41467-024-49754-4} {\bibfield  {journal} {\bibinfo
  {journal} {Nature Commun.}\ }\textbf {\bibinfo {volume} {15}},\ \bibinfo
  {pages} {5466} (\bibinfo {year} {2024})}\BibitemShut {NoStop}%
\bibitem [{\citenamefont {Jagla}(1998)}]{Jagla98}%
  \BibitemOpen
  \bibfield  {author} {\bibinfo {author} {\bibfnamefont {E.~A.}\ \bibnamefont
  {Jagla}},\ }\bibfield  {title} {\bibinfo {title} {Phase behavior of a system
  of particles with core collapse},\ }\href
  {https://doi.org/10.1103/PhysRevE.58.1478} {\bibfield  {journal} {\bibinfo
  {journal} {Phys. Rev. E}\ }\textbf {\bibinfo {volume} {58}},\ \bibinfo
  {pages} {1478} (\bibinfo {year} {1998})}\BibitemShut {NoStop}%
\bibitem [{\citenamefont {Malescio}\ and\ \citenamefont
  {Pellicane}(2003)}]{Malescio03}%
  \BibitemOpen
  \bibfield  {author} {\bibinfo {author} {\bibfnamefont {G.}~\bibnamefont
  {Malescio}}\ and\ \bibinfo {author} {\bibfnamefont {G.}~\bibnamefont
  {Pellicane}},\ }\bibfield  {title} {\bibinfo {title} {Stripe phases from
  isotropic repulsive interactions},\ }\href {https://doi.org/10.1038/nmat820}
  {\bibfield  {journal} {\bibinfo  {journal} {Nature Mater.}\ }\textbf
  {\bibinfo {volume} {2}},\ \bibinfo {pages} {97} (\bibinfo {year}
  {2003})}\BibitemShut {NoStop}%
\bibitem [{\citenamefont {Glaser}\ \emph {et~al.}(2007)\citenamefont {Glaser},
  \citenamefont {Grason}, \citenamefont {Kamien}, \citenamefont {Kosmrlj},
  \citenamefont {Santangelo},\ and\ \citenamefont {Ziherl}}]{Glaser07}%
  \BibitemOpen
  \bibfield  {author} {\bibinfo {author} {\bibfnamefont {M.~A.}\ \bibnamefont
  {Glaser}}, \bibinfo {author} {\bibfnamefont {G.~M.}\ \bibnamefont {Grason}},
  \bibinfo {author} {\bibfnamefont {R.~D.}\ \bibnamefont {Kamien}}, \bibinfo
  {author} {\bibfnamefont {A.}~\bibnamefont {Kosmrlj}}, \bibinfo {author}
  {\bibfnamefont {C.~D.}\ \bibnamefont {Santangelo}},\ and\ \bibinfo {author}
  {\bibfnamefont {P.}~\bibnamefont {Ziherl}},\ }\bibfield  {title} {\bibinfo
  {title} {Soft spheres make more mesophases},\ }\href
  {https://doi.org/10.1209/0295-5075/78/46004} {\bibfield  {journal} {\bibinfo
  {journal} {EPL}\ }\textbf {\bibinfo {volume} {78}},\ \bibinfo {pages} {46004}
  (\bibinfo {year} {2007})}\BibitemShut {NoStop}%
\bibitem [{\citenamefont {Costa~Campos}\ \emph {et~al.}(2013)\citenamefont
  {Costa~Campos}, \citenamefont {Apolinario},\ and\ \citenamefont
  {L\"owen}}]{CostaCampos13}%
  \BibitemOpen
  \bibfield  {author} {\bibinfo {author} {\bibfnamefont {L.~Q.}\ \bibnamefont
  {Costa~Campos}}, \bibinfo {author} {\bibfnamefont {S.~W.~S.}\ \bibnamefont
  {Apolinario}},\ and\ \bibinfo {author} {\bibfnamefont {H.}~\bibnamefont
  {L\"owen}},\ }\bibfield  {title} {\bibinfo {title} {Structural ordering of
  trapped colloids with competing interactions},\ }\href
  {https://doi.org/10.1103/PhysRevE.88.042313} {\bibfield  {journal} {\bibinfo
  {journal} {Phys. Rev. E}\ }\textbf {\bibinfo {volume} {88}},\ \bibinfo
  {pages} {042313} (\bibinfo {year} {2013})}\BibitemShut {NoStop}%
\bibitem [{\citenamefont {Al~Harraq}\ \emph {et~al.}(2022)\citenamefont
  {Al~Harraq}, \citenamefont {Hymel}, \citenamefont {Lin}, \citenamefont
  {Truskett},\ and\ \citenamefont {Bharti}}]{AlHarraq22}%
  \BibitemOpen
  \bibfield  {author} {\bibinfo {author} {\bibfnamefont {A.}~\bibnamefont
  {Al~Harraq}}, \bibinfo {author} {\bibfnamefont {A.~A.}\ \bibnamefont
  {Hymel}}, \bibinfo {author} {\bibfnamefont {E.}~\bibnamefont {Lin}}, \bibinfo
  {author} {\bibfnamefont {T.~M.}\ \bibnamefont {Truskett}},\ and\ \bibinfo
  {author} {\bibfnamefont {B.}~\bibnamefont {Bharti}},\ }\bibfield  {title}
  {\bibinfo {title} {Dual nature of magnetic nanoparticle dispersions enables
  control over short-range attraction and long-range repulsion interactions},\
  }\href {https://doi.org/10.1038/s42004-022-00687-3} {\bibfield  {journal}
  {\bibinfo  {journal} {Commun. Chem.}\ }\textbf {\bibinfo {volume} {5}},\
  \bibinfo {pages} {72} (\bibinfo {year} {2022})}\BibitemShut {NoStop}%
\bibitem [{\citenamefont {Fogler}\ \emph {et~al.}(1996)\citenamefont {Fogler},
  \citenamefont {Koulakov},\ and\ \citenamefont {Shklovskii}}]{Fogler96}%
  \BibitemOpen
  \bibfield  {author} {\bibinfo {author} {\bibfnamefont {M.~M.}\ \bibnamefont
  {Fogler}}, \bibinfo {author} {\bibfnamefont {A.~A.}\ \bibnamefont
  {Koulakov}},\ and\ \bibinfo {author} {\bibfnamefont {B.~I.}\ \bibnamefont
  {Shklovskii}},\ }\bibfield  {title} {\bibinfo {title} {Ground state of a
  two-dimensional electron liquid in a weak magnetic field},\ }\href
  {https://doi.org/10.1103/PhysRevB.54.1853} {\bibfield  {journal} {\bibinfo
  {journal} {Phys. Rev. B}\ }\textbf {\bibinfo {volume} {54}},\ \bibinfo
  {pages} {1853} (\bibinfo {year} {1996})}\BibitemShut {NoStop}%
\bibitem [{\citenamefont {Moessner}\ and\ \citenamefont
  {Chalker}(1996)}]{Moessner96}%
  \BibitemOpen
  \bibfield  {author} {\bibinfo {author} {\bibfnamefont {R.}~\bibnamefont
  {Moessner}}\ and\ \bibinfo {author} {\bibfnamefont {J.~T.}\ \bibnamefont
  {Chalker}},\ }\bibfield  {title} {\bibinfo {title} {Exact results for
  interacting electrons in high {L}andau levels},\ }\href
  {https://doi.org/10.1103/PhysRevB.54.5006} {\bibfield  {journal} {\bibinfo
  {journal} {Phys. Rev. B}\ }\textbf {\bibinfo {volume} {54}},\ \bibinfo
  {pages} {5006} (\bibinfo {year} {1996})}\BibitemShut {NoStop}%
\bibitem [{\citenamefont {Cooper}\ \emph {et~al.}(1999)\citenamefont {Cooper},
  \citenamefont {Lilly}, \citenamefont {Eisenstein}, \citenamefont {Pfeiffer},\
  and\ \citenamefont {West}}]{Cooper99}%
  \BibitemOpen
  \bibfield  {author} {\bibinfo {author} {\bibfnamefont {K.~B.}\ \bibnamefont
  {Cooper}}, \bibinfo {author} {\bibfnamefont {M.~P.}\ \bibnamefont {Lilly}},
  \bibinfo {author} {\bibfnamefont {J.~P.}\ \bibnamefont {Eisenstein}},
  \bibinfo {author} {\bibfnamefont {L.~N.}\ \bibnamefont {Pfeiffer}},\ and\
  \bibinfo {author} {\bibfnamefont {K.~W.}\ \bibnamefont {West}},\ }\bibfield
  {title} {\bibinfo {title} {Insulating phases of two-dimensional electrons in
  high {L}andau levels: Observation of sharp thresholds to conduction},\ }\href
  {https://doi.org/10.1103/PhysRevB.60.R11285} {\bibfield  {journal} {\bibinfo
  {journal} {Phys. Rev. B}\ }\textbf {\bibinfo {volume} {60}},\ \bibinfo
  {pages} {R11285} (\bibinfo {year} {1999})}\BibitemShut {NoStop}%
\bibitem [{\citenamefont {Fradkin}\ and\ \citenamefont
  {Kivelson}(1999)}]{Fradkin99}%
  \BibitemOpen
  \bibfield  {author} {\bibinfo {author} {\bibfnamefont {E.}~\bibnamefont
  {Fradkin}}\ and\ \bibinfo {author} {\bibfnamefont {S.~A.}\ \bibnamefont
  {Kivelson}},\ }\bibfield  {title} {\bibinfo {title} {Liquid-crystal phases of
  quantum {Hall} systems},\ }\href {https://doi.org/10.1103/PhysRevB.59.8065}
  {\bibfield  {journal} {\bibinfo  {journal} {Phys. Rev. B}\ }\textbf {\bibinfo
  {volume} {59}},\ \bibinfo {pages} {8065} (\bibinfo {year}
  {1999})}\BibitemShut {NoStop}%
\bibitem [{\citenamefont {G\"ores}\ \emph {et~al.}(2007)\citenamefont
  {G\"ores}, \citenamefont {Gamez}, \citenamefont {Smet}, \citenamefont
  {Pfeiffer}, \citenamefont {West}, \citenamefont {Yacoby}, \citenamefont
  {Umansky},\ and\ \citenamefont {von Klitzing}}]{Gores07}%
  \BibitemOpen
  \bibfield  {author} {\bibinfo {author} {\bibfnamefont {J.}~\bibnamefont
  {G\"ores}}, \bibinfo {author} {\bibfnamefont {G.}~\bibnamefont {Gamez}},
  \bibinfo {author} {\bibfnamefont {J.~H.}\ \bibnamefont {Smet}}, \bibinfo
  {author} {\bibfnamefont {L.}~\bibnamefont {Pfeiffer}}, \bibinfo {author}
  {\bibfnamefont {K.}~\bibnamefont {West}}, \bibinfo {author} {\bibfnamefont
  {A.}~\bibnamefont {Yacoby}}, \bibinfo {author} {\bibfnamefont
  {V.}~\bibnamefont {Umansky}},\ and\ \bibinfo {author} {\bibfnamefont
  {K.}~\bibnamefont {von Klitzing}},\ }\bibfield  {title} {\bibinfo {title}
  {Current-induced anisotropy and reordering of the electron liquid-crystal
  phases in a two-dimensional electron system},\ }\href
  {https://doi.org/10.1103/PhysRevLett.99.246402} {\bibfield  {journal}
  {\bibinfo  {journal} {Phys. Rev. Lett.}\ }\textbf {\bibinfo {volume} {99}},\
  \bibinfo {pages} {246402} (\bibinfo {year} {2007})}\BibitemShut {NoStop}%
\bibitem [{\citenamefont {Friess}\ \emph {et~al.}(2018)\citenamefont {Friess},
  \citenamefont {Umansky}, \citenamefont {von Klitzing},\ and\ \citenamefont
  {Smet}}]{Friess18}%
  \BibitemOpen
  \bibfield  {author} {\bibinfo {author} {\bibfnamefont {B.}~\bibnamefont
  {Friess}}, \bibinfo {author} {\bibfnamefont {V.}~\bibnamefont {Umansky}},
  \bibinfo {author} {\bibfnamefont {K.}~\bibnamefont {von Klitzing}},\ and\
  \bibinfo {author} {\bibfnamefont {J.~H.}\ \bibnamefont {Smet}},\ }\bibfield
  {title} {\bibinfo {title} {Current flow in the bubble and stripe phases},\
  }\href {https://doi.org/10.1103/PhysRevLett.120.137603} {\bibfield  {journal}
  {\bibinfo  {journal} {Phys. Rev. Lett.}\ }\textbf {\bibinfo {volume} {120}},\
  \bibinfo {pages} {137603} (\bibinfo {year} {2018})}\BibitemShut {NoStop}%
\bibitem [{\citenamefont {Xu}\ \emph {et~al.}(2021)\citenamefont {Xu},
  \citenamefont {Tang}, \citenamefont {Wang}, \citenamefont {Xu}, \citenamefont
  {Fang},\ and\ \citenamefont {Gu}}]{Xu21}%
  \BibitemOpen
  \bibfield  {author} {\bibinfo {author} {\bibfnamefont {X.~B.}\ \bibnamefont
  {Xu}}, \bibinfo {author} {\bibfnamefont {T.}~\bibnamefont {Tang}}, \bibinfo
  {author} {\bibfnamefont {Z.~H.}\ \bibnamefont {Wang}}, \bibinfo {author}
  {\bibfnamefont {X.~N.}\ \bibnamefont {Xu}}, \bibinfo {author} {\bibfnamefont
  {G.~Y.}\ \bibnamefont {Fang}},\ and\ \bibinfo {author} {\bibfnamefont
  {M.}~\bibnamefont {Gu}},\ }\bibfield  {title} {\bibinfo {title}
  {Nonequilibrium pattern formation in circularly confined two-dimensional
  systems with competing interactions},\ }\href
  {https://doi.org/10.1103/PhysRevE.103.012604} {\bibfield  {journal} {\bibinfo
   {journal} {Phys. Rev. E}\ }\textbf {\bibinfo {volume} {103}},\ \bibinfo
  {pages} {012604} (\bibinfo {year} {2021})}\BibitemShut {NoStop}%
\bibitem [{\citenamefont {Komendov\'a}\ \emph {et~al.}(2012)\citenamefont
  {Komendov\'a}, \citenamefont {Chen}, \citenamefont {Shanenko}, \citenamefont
  {Milo\ifmmode \check{s}\else \v{s}\fi{}evi\ifmmode~\acute{c}\else
  \'{c}\fi{}},\ and\ \citenamefont {Peeters}}]{Komendova12}%
  \BibitemOpen
  \bibfield  {author} {\bibinfo {author} {\bibfnamefont {L.}~\bibnamefont
  {Komendov\'a}}, \bibinfo {author} {\bibfnamefont {Y.}~\bibnamefont {Chen}},
  \bibinfo {author} {\bibfnamefont {A.~A.}\ \bibnamefont {Shanenko}}, \bibinfo
  {author} {\bibfnamefont {M.~V.}\ \bibnamefont {Milo\ifmmode \check{s}\else
  \v{s}\fi{}evi\ifmmode~\acute{c}\else \'{c}\fi{}}},\ and\ \bibinfo {author}
  {\bibfnamefont {F.~M.}\ \bibnamefont {Peeters}},\ }\bibfield  {title}
  {\bibinfo {title} {Two-band superconductors: Hidden criticality deep in the
  superconducting state},\ }\href
  {https://doi.org/10.1103/PhysRevLett.108.207002} {\bibfield  {journal}
  {\bibinfo  {journal} {Phys. Rev. Lett.}\ }\textbf {\bibinfo {volume} {108}},\
  \bibinfo {pages} {207002} (\bibinfo {year} {2012})}\BibitemShut {NoStop}%
\bibitem [{\citenamefont {Varney}\ \emph {et~al.}(2013)\citenamefont {Varney},
  \citenamefont {Sellin}, \citenamefont {Wang}, \citenamefont {Fangohr},\ and\
  \citenamefont {Babaev}}]{Varney13}%
  \BibitemOpen
  \bibfield  {author} {\bibinfo {author} {\bibfnamefont {C.~N.}\ \bibnamefont
  {Varney}}, \bibinfo {author} {\bibfnamefont {K.~A.~H.}\ \bibnamefont
  {Sellin}}, \bibinfo {author} {\bibfnamefont {Q.-Z.}\ \bibnamefont {Wang}},
  \bibinfo {author} {\bibfnamefont {H.}~\bibnamefont {Fangohr}},\ and\ \bibinfo
  {author} {\bibfnamefont {E.}~\bibnamefont {Babaev}},\ }\bibfield  {title}
  {\bibinfo {title} {Hierarchical structure foramtion in layered
  superconducting systems with multi-scale inter-vortex interactions},\ }\href
  {https://doi.org/10.1088/0953-8984/25/41/415702} {\bibfield  {journal}
  {\bibinfo  {journal} {J. Phys.: Condens. Matter}\ }\textbf {\bibinfo {volume}
  {25}},\ \bibinfo {pages} {415702} (\bibinfo {year} {2013})}\BibitemShut
  {NoStop}%
\bibitem [{\citenamefont {Sellin}\ and\ \citenamefont
  {Babaev}(2013)}]{Sellin13}%
  \BibitemOpen
  \bibfield  {author} {\bibinfo {author} {\bibfnamefont {K.~A.~H.}\
  \bibnamefont {Sellin}}\ and\ \bibinfo {author} {\bibfnamefont
  {E.}~\bibnamefont {Babaev}},\ }\bibfield  {title} {\bibinfo {title} {Stripe,
  gossamer, and glassy phases in systems with strong nonpairwise
  interactions},\ }\href {https://doi.org/10.1103/PhysRevE.88.042305}
  {\bibfield  {journal} {\bibinfo  {journal} {Phys. Rev. E}\ }\textbf {\bibinfo
  {volume} {88}},\ \bibinfo {pages} {042305} (\bibinfo {year}
  {2013})}\BibitemShut {NoStop}%
\bibitem [{\citenamefont {Brems}\ \emph {et~al.}(2022)\citenamefont {Brems},
  \citenamefont {M{\" u}hlbauer}, \citenamefont {C{\' o}rdoba-Camacho},
  \citenamefont {Shanenko}, \citenamefont {Vagov}, \citenamefont {Aguiar},\
  and\ \citenamefont {Cubitt}}]{Brems22}%
  \BibitemOpen
  \bibfield  {author} {\bibinfo {author} {\bibfnamefont {X.~S.}\ \bibnamefont
  {Brems}}, \bibinfo {author} {\bibfnamefont {S.}~\bibnamefont {M{\"
  u}hlbauer}}, \bibinfo {author} {\bibfnamefont {W.~Y.}\ \bibnamefont {C{\'
  o}rdoba-Camacho}}, \bibinfo {author} {\bibfnamefont {A.~A.}\ \bibnamefont
  {Shanenko}}, \bibinfo {author} {\bibfnamefont {A.}~\bibnamefont {Vagov}},
  \bibinfo {author} {\bibfnamefont {J.~A.}\ \bibnamefont {Aguiar}},\ and\
  \bibinfo {author} {\bibfnamefont {R.}~\bibnamefont {Cubitt}},\ }\bibfield
  {title} {\bibinfo {title} {Current-induced self-organisation of mixed
  superconducting states},\ }\href {https://doi.org/10.1088/1361-6668/ac455e}
  {\bibfield  {journal} {\bibinfo  {journal} {Supercond. Sci. Technol.}\
  }\textbf {\bibinfo {volume} {35}},\ \bibinfo {pages} {035003} (\bibinfo
  {year} {2022})}\BibitemShut {NoStop}%
\bibitem [{\citenamefont {McDermott}\ \emph {et~al.}(2016)\citenamefont
  {McDermott}, \citenamefont {Reichhardt},\ and\ \citenamefont
  {Reichhardt}}]{McDermott16}%
  \BibitemOpen
  \bibfield  {author} {\bibinfo {author} {\bibfnamefont {D.}~\bibnamefont
  {McDermott}}, \bibinfo {author} {\bibfnamefont {C.~J.~O.}\ \bibnamefont
  {Reichhardt}},\ and\ \bibinfo {author} {\bibfnamefont {C.}~\bibnamefont
  {Reichhardt}},\ }\bibfield  {title} {\bibinfo {title} {Structural transitions
  and hysteresis in clump- and stripe-forming systems under dynamic
  compression},\ }\href {https://doi.org/10.1039/C6SM01939K} {\bibfield
  {journal} {\bibinfo  {journal} {Soft Matter}\ }\textbf {\bibinfo {volume}
  {12}},\ \bibinfo {pages} {9549} (\bibinfo {year} {2016})}\BibitemShut
  {NoStop}%
\bibitem [{\citenamefont {Reichhardt}\ and\ \citenamefont
  {Reichhardt}(2024{\natexlab{a}})}]{Reichhardt24}%
  \BibitemOpen
  \bibfield  {author} {\bibinfo {author} {\bibfnamefont {C.}~\bibnamefont
  {Reichhardt}}\ and\ \bibinfo {author} {\bibfnamefont {C.~J.~O.}\ \bibnamefont
  {Reichhardt}},\ }\bibfield  {title} {\bibinfo {title} {Peak effect and
  dynamics of stripe- and pattern-forming systems on a periodic one-dimensional
  substrate},\ }\href {https://doi.org/10.1103/PhysRevE.109.054606} {\bibfield
  {journal} {\bibinfo  {journal} {Phys. Rev. E}\ }\textbf {\bibinfo {volume}
  {109}},\ \bibinfo {pages} {054606} (\bibinfo {year}
  {2024}{\natexlab{a}})}\BibitemShut {NoStop}%
\bibitem [{\citenamefont {Reichhardt}\ and\ \citenamefont
  {Reichhardt}(2024{\natexlab{b}})}]{Reichhardt24a}%
  \BibitemOpen
  \bibfield  {author} {\bibinfo {author} {\bibfnamefont {C.}~\bibnamefont
  {Reichhardt}}\ and\ \bibinfo {author} {\bibfnamefont {C.~J.~O.}\ \bibnamefont
  {Reichhardt}},\ }\bibfield  {title} {\bibinfo {title} {Stripe and bubble
  ratchets on asymmetric substrates},\ }\href
  {https://doi.org/10.1103/PhysRevResearch.6.043290} {\bibfield  {journal}
  {\bibinfo  {journal} {Phys. Rev. Res.}\ }\textbf {\bibinfo {volume} {6}},\
  \bibinfo {pages} {043290} (\bibinfo {year} {2024}{\natexlab{b}})}\BibitemShut
  {NoStop}%
\bibitem [{\citenamefont {Reichhardt}\ and\ \citenamefont
  {Reichhardt}(2014)}]{Reichhardt14}%
  \BibitemOpen
  \bibfield  {author} {\bibinfo {author} {\bibfnamefont {C.}~\bibnamefont
  {Reichhardt}}\ and\ \bibinfo {author} {\bibfnamefont {C.~J.~O.}\ \bibnamefont
  {Reichhardt}},\ }\bibfield  {title} {\bibinfo {title} {Aspects of jamming in
  two-dimensional athermal frictionless systems},\ }\href
  {https://doi.org/10.1039/c3sm53154f} {\bibfield  {journal} {\bibinfo
  {journal} {Soft Matter}\ }\textbf {\bibinfo {volume} {10}},\ \bibinfo {pages}
  {2932} (\bibinfo {year} {2014})}\BibitemShut {NoStop}%
\bibitem [{\citenamefont {Dahmen}\ \emph {et~al.}(2011)\citenamefont {Dahmen},
  \citenamefont {Ben-Zion},\ and\ \citenamefont {Uhl}}]{Dahmen11}%
  \BibitemOpen
  \bibfield  {author} {\bibinfo {author} {\bibfnamefont {K.~A.}\ \bibnamefont
  {Dahmen}}, \bibinfo {author} {\bibfnamefont {Y.}~\bibnamefont {Ben-Zion}},\
  and\ \bibinfo {author} {\bibfnamefont {J.~T.}\ \bibnamefont {Uhl}},\
  }\bibfield  {title} {\bibinfo {title} {A simple analytic theory for the
  statistics of avalanches in sheared granular materials},\ }\href
  {https://doi.org/10.1038/NPHYS1957} {\bibfield  {journal} {\bibinfo
  {journal} {Nature Phys.}\ }\textbf {\bibinfo {volume} {7}},\ \bibinfo {pages}
  {554} (\bibinfo {year} {2011})}\BibitemShut {NoStop}%
\bibitem [{\citenamefont {Reichhardt}\ and\ \citenamefont
  {Reichhardt}(2017)}]{Reichhardt17}%
  \BibitemOpen
  \bibfield  {author} {\bibinfo {author} {\bibfnamefont {C.}~\bibnamefont
  {Reichhardt}}\ and\ \bibinfo {author} {\bibfnamefont {C.~J.~O.}\ \bibnamefont
  {Reichhardt}},\ }\bibfield  {title} {\bibinfo {title} {Depinning and
  nonequilibrium dynamic phases of particle assemblies driven over random and
  ordered substrates: a review},\ }\href
  {https://doi.org/10.1088/1361-6633/80/2/026501} {\bibfield  {journal}
  {\bibinfo  {journal} {Rep. Prog. Phys.}\ }\textbf {\bibinfo {volume} {80}},\
  \bibinfo {pages} {026501} (\bibinfo {year} {2017})}\BibitemShut {NoStop}%
\bibitem [{\citenamefont {Reichhardt}\ \emph {et~al.}(2025)\citenamefont
  {Reichhardt}, \citenamefont {McDermott},\ and\ \citenamefont
  {Reichhardt}}]{Reichhardt25}%
  \BibitemOpen
  \bibfield  {author} {\bibinfo {author} {\bibfnamefont {C.~J.~O.}\
  \bibnamefont {Reichhardt}}, \bibinfo {author} {\bibfnamefont
  {D.}~\bibnamefont {McDermott}},\ and\ \bibinfo {author} {\bibfnamefont
  {C.}~\bibnamefont {Reichhardt}},\ }\bibfield  {title} {\bibinfo {title}
  {Using principal component analysis to distinguish different dynamic phases
  in superconducting vortex matter},\ }\href
  {https://doi.org/10.1103/PhysRevB.111.104508} {\bibfield  {journal} {\bibinfo
   {journal} {Phys. Rev. B}\ }\textbf {\bibinfo {volume} {111}},\ \bibinfo
  {pages} {104508} (\bibinfo {year} {2025})}\BibitemShut {NoStop}%
\end{thebibliography}%

\end{document}